\documentclass[]{IEEEtran}
\usepackage{cite}
\usepackage{amsmath,amssymb,amsfonts}
\usepackage{algorithmic}
\usepackage{graphicx}
\usepackage{textcomp}
\usepackage{xcolor}

\ifCLASSOPTIONcompsoc
    \usepackage[caption=false, font=normalsize, labelfont=sf, textfont=sf]{subfig}
\else
\usepackage[caption=false, font=footnotesize]{subfig}
\fi

\usepackage{url}								
\usepackage{siunitx}							
\usepackage{booktabs}							
\usepackage{tabularx}
\newcolumntype{Y}{>{\centering\arraybackslash}X} 

\usepackage{makecell}							
\usepackage{multirow}							
\usepackage[display-foreign=false]{acro}			
\usepackage{gensymb}
\acsetup{single}								

\usepackage{balance}							
\usepackage{tikz}								

\usepackage{soul}
\usepackage[most]{tcolorbox}

\DeclareAcronym{lfm}{
	short = LFM,
	long = linear frequency modulation
}
\DeclareAcronym{prf}{
	short = PRF,
	long = pulse repetition frequency
}
\DeclareAcronym{pri}{
	short = PRI,
	long = pulse repetition interval
}
\DeclareAcronym{fmcw}{
	short = FMCW,
	long = frequency modulated continuous-wave
}
\DeclareAcronym{lfmcw}{
	short = LFMCW,
	long = linear frequency modulated continuous-wave
}
\DeclareAcronym{cw}{
	short = CW,
	long = continuous-wave
}
\DeclareAcronym{dbf}{
	short = DBF,
	long = digital beamforming
}
\DeclareAcronym{sar}{
	short = SAR,
	long = synthetic aperture radar
}
\DeclareAcronym{psr}{
	short = PSR,
	long = point scatterer response
}
\DeclareAcronym{rcs}{
	short = RCS,
	long = radar cross-section
}
\DeclareAcronym{crlb}{
	short = CRLB,
	long = Cramer-Rao lower bound
}
\DeclareAcronym{dof}{
	short = DoF,
	long = degree of freedom
}
\DeclareAcronym{snr}{
	short = SNR,
	long = signal-to-noise ratio
}
\DeclareAcronym{sinr}{
	short = SINR,
	long = signal-to-interference-plus-noise ratio
}
\DeclareAcronym{fft}{
	short = FFT,
	long = fast Fourier transform,
}
\DeclareAcronym{ifft}{
	short = IFFT,
	long = inverse \ac{fft},
}
\DeclareAcronym{ift}{
	short = IFT,
	long = inverse Fourier transform,
}
\DeclareAcronym{rmse}{
	short = RMSE,
	long = root-mean-square error
}
\DeclareAcronym{psd}{
	short = PSD,
	long = power spectral density
}
\DeclareAcronym{rca}{
	short = RCA,
	long = range of closest approach
}
\DeclareAcronym{rda}{
	short = RDA,
	long = Range-Doppler Algorithm
}
\DeclareAcronym{rma}{
	short = RMA,
	long = Range Migration Algorithm
}
\DeclareAcronym{pfa}{
	short = PFA,
	long = Polar Formatting Algorithm
}
\DeclareAcronym{bpa}{
	short = BPA,
	long = Backprojection Algorithm
}
\DeclareAcronym{rvp}{
	short = RVP,
	long = residual video phase
}
\DeclareAcronym{jrc}{
	short = JRC,
	long = joint radar-communications
}
\DeclareAcronym{doa}{
	short = DOA,
	long = direction of arrival
}
\DeclareAcronym{hci}{
	short = HCI,
	long = human-computer interaction
}
\DeclareAcronym{its}{
	short = ITS,
	long = intelligent transportation systems
}
\DeclareAcronym{rtk}{
	short = RTK,
	long = real-time kinematic
}
\DeclareAcronym{eirp}{
	short = EIRP,
	long = effective isotropic radiated power
}
\DeclareAcronym{gnss}{
	short = GNSS,
	long = global navigation satellite system
}
\DeclareAcronym{imu}{
	short = IMU,
	long = inertial measurement unit
}

\DeclareAcronym{ofdm}{
	short = OFDM,
	long = orthogonal frequency division multiplexing
}

\DeclareAcronym{pll}{
	short = PLL,
	long = phase-locked loop
}
\DeclareAcronym{vco}{
	short = VCO,
	long = voltage-controlled oscillator
}
\DeclareAcronym{lna}{
	short = LNA,
	long = low-noise amplifier
}
\DeclareAcronym{if}{
	short = IF,
	long = intermediate frequency
}
\DeclareAcronym{cots}{
	short = COTS,
	long = commercial off-the-shelf
}
\DeclareAcronym{adc}{
	short = ADC,
	long = analog to digital converter
}
\DeclareAcronym{lo}{
	short = LO,
	long = local oscillator
}
\DeclareAcronym{pcb}{
	short = PCB,
	long = printed circuit board
}
\DeclareAcronym{mimo}{
	short = MIMO,
	long = multiple-input multiple-output
}
\DeclareAcronym{simo}{
	short = SIMO,
	long = single-input multiple-output
}
\DeclareAcronym{mmic}{
	short = MMIC,
	long = monolithic microwave integrated circuit
}
\DeclareAcronym{daq}{
	short = DAQ,
	long = data acquisition
}
\DeclareAcronym{ic}{
	short = IC,
	long = integrated circuit
}
\DeclareAcronym{pa}{
	short = PA,
	long = power amplifier
}

\DeclareAcronym{ti}{
	short = TI,
	long = Texas Instruments
}
\DeclareAcronym{adi}{
	short = ADI,
	long = Analog Devices
}

\DeclareAcronym{roi}{
	short = ROI,
	long = region of interest,
	long-plural-form = regions of interest
}
\DeclareAcronym{v2x}{
	short = V2X,
	long = vehicle-to-everything
}
\DeclareAcronym{av}{
	short = AV,
	long = automated vehicle
}
\DeclareAcronym{cors}{
	short = CORS,
	long = continuously operating reference station
}
\DeclareAcronym{mdot}{
	short = MDOT,
	long = Michigan Department of Transportation
}
\DeclareAcronym{moco}{
	short = MOCO,
	long = motion compensation
}

\newcommand{\hlb}[1]{#1}

\newcolumntype{Y}{>{\centering\arraybackslash}X}

\newtcolorbox{hlbox}[1][]{%
  colback=white,
  float=htb,
  boxsep=0pt,
  #1%
}

\def\BibTeX{{\rm B\kern-.05em{\sc i\kern-.025em b}\kern-.08em
    T\kern-.1667em\lower.7ex\hbox{E}\kern-.125emX}}
\begin{document}

\title{A \hlb{C}-Band Fully Polarimetric Automotive Synthetic Aperture Radar}

\author{ Jason M. Merlo,~\IEEEmembership{Graduate~Student~Member,~IEEE,} and Jeffrey A. Nanzer,~\IEEEmembership{Senior Member,~IEEE}%
	\thanks{Copyright \copyright 2021 IEEE. Personal use of this material is permitted. However, permission to use this material for any other purposes must be obtained from the IEEE by sending a request to pubs-permissions@ieee.org.}
	\thanks{	
	The authors are with the Department of Electrical and Computer Engineering, Michigan State University, East Lansing, MI 48824 USA (email: merlojas@msu.edu, nanzer@msu.edu).}
}

\maketitle

\begin{abstract}
Due to the rapid increase in \SI{76}{\giga\hertz} automotive spectrum use in recent years, wireless interference is becoming a legitimate area of concern. However, the recent rise in interest of \acp{av} has also spurred new growth and adoption of low frequency \ac{v2x} communications in and around the \SI{5.8}{\giga\hertz} unlicensed bands, opening the possibility for communications spectrum reuse in the form of \acf{jrc}. In this work, we present a low frequency \SI{5.9}{\giga\hertz} side-looking polarimetric \ac{sar} for automotive use, utilizing a ranging waveform in a common low frequency \ac{v2x} communications band. A synthetic aperture technique is employed to address the angular resolution concerns commonly associated with radars at lower frequencies. Three side-looking fully polarimetric \ac{sar} images in various urban scenes are presented and discussed to highlight the unique opportunities for landmark inference afforded through measurement of co- and cross-polarized scattering. 
\end{abstract}


\acresetall 

\section{Introduction}


\IEEEPARstart{A}{utomated} vehicles have experienced a rapid growth in recent years, and with this, a flurry of new technologies have arisen to satisfy the growing demand for a high resolution representation of the vehicle's surroundings. Situational sensing will be necessary for many driving tasks, especially as the degree of automation progresses towards levels 4 and 5 \cite{J3016_201806}. Typically, these high resolution maps are generated by means of one or more lidars in conjunction with cameras, short-range ultrasonic sonar, and \ac{cots} automotive radar units with angular resolutions on the order of several degrees; these data may then be fused to produce occupancy grids for object detection and navigation, as well as for localization and mapping. While optical means of sensing can provide high spatial resolution, camera imagery often degrades in low-light situations and cannot intrinsically determine depth in monocular configurations; lidar can also be inhibited by high attenuation in inclement weather scenarios such as rain, snow, and dense fog \cite{jokela2019testing}. Sonar is helpful for low-speed, short-range object avoidance, however it's spatial resolution and maximum range are typically restrictive for more dynamic driving scenarios. Similar to lidar, sonar is also affected by increased attenuation in adverse weather conditions such as snow and heavy rain. Radar, due to its relatively long wavelengths, is more robust to precipitation and other atmospheric attenuation, and, at lower frequencies, it is even able to propagate through some vegetation, enabling it to operate at longer ranges than other modalities. In addition, radar is commonly used to measure relative velocity of objects in a vehicle's surrounding using the Doppler effect. Because of these attributes, radar is often utilized to augment the suite of sensors on \acp{av}, providing an improved spatial awareness in all weather and lighting conditions.  The common drawback to using radar is typically in its angular resolution as it is inversely proportional to the physical size of the aperture and its wavelength, commonly necessitating large arrays to achieve sub-degree angular resolution. Due to the desire to create compact arrays on automotive vehicles, the trend in automotive radar has been to increase the carrier frequency to improve the range and angular resolutions obtainable while maintaining compact antenna and array sizes. However, current trends in vehicular communications, motivated by recent interest in connected and \acp{av}, are opening new low frequency spectrums for use on vehicles, while simultaneously, the proliferation of the \SI{76}{\giga\hertz} spectrum commonly used in automotive radar is causing increased noise and interference on dedicated ranging frequencies. This offers a unique opportunity to investigate spectrum reuse in the \ac{v2x} communications bands in and around the \SI{5.8}{\giga\hertz} unlicensed spectrum. While spectrum reuse is currently garnering significant interest  for \acf{jrc} applications \cite{moghaddasi2013improved, reichardt2012demonstrating, ellison2020high, kumari2017ieee, liu2020joint, duggal2020doppler}, many of these techniques focus on higher frequencies as the use of lower carrier frequencies typically necessitates physically large arrays and affords poor spatial resolution making discriminating road hazards difficult; consequently, these frequencies have typically been avoided for use in automotive radars.


To address the challenge of angular resolution in automotive radar, a handful of techniques have been investigated and implemented over the last few decades.  Initially, mechanically steered array techniques were implemented to steer a narrow transmit beam across the scene to obtain a range profile in a given direction, later the use of digitally scanned arrays enabled rapid beam steering and \ac{dbf} on receive was implemented to allow \ac{doa} estimation from all directions simultaneously \cite{waldschmidt2021automotive}. Shortly thereafter, \acf{mimo} techniques were introduced which achieve improved angular resolution by creating a virtual array of antennas which is greater than the sum of physical antennas through the use of orthogonal transmit waveforms\cite{waldschmidt2021automotive,li2008mimo,pfeffer2013fmcw}; these techniques are still heavily relied upon in automotive radar today \cite{bilik2016automotive,roos2019compressed,sun2020mimo, farhadi2021automotive}. However, current trends are moving towards utilizing distributed networks of radars on vehicle to achieve greater aperture sizes, enabling finer angular resolution. One densely filled aperture technique was presented in \cite{bialer2020performance} which used a \SI{1}{\meter} wide array consisting of 15 coherent \ac{mimo} sub-arrays and a total of 180 and 240 transmit and receive antennas, respectively, achieved an angular resolution on the order of a tenth of a degree, however the size of the array presents a manufacturing challenge for large scale automotive environments. Another physically large aperture approach \cite{gottinger2021coherent} utilized a sparsely filled aperture consisting of two coherent, distributed \ac{mimo} sub-arrays spaced \SI{0.86}{\meter} apart achieving \ang{1.7} of angular resolution in azimuth and \ang{3} in elevation.  In the last decade, others have explored the use of synthetic aperture techniques to produce even larger apertures by moving a single element, or sub-array along the track of the vehicle to produce a single large synthetic aperture. \hlb{Early works focused primarily on the use of \ac{sar} for occupancy grids in scenarios such as parking spot occupancy identification, starting with \SI{24}{\giga\hertz}\cite{wu2009automotive, wu2010novel} before moving to the \SI{76}{\giga\hertz} band to utilize the improved ranging resolution afforded by the larger bandwidths typically achievable at these frequencies and improved crossrange resolution, which is proportional to the wavelength \cite{iqbal2015sar, feger2017experimental, harrer2017synthetic, laribi2018performance, steiner2020millimeter, farhadi2021automotive, iqbal2021imaging, tagliaferri2021navigation, iqbal2021realistic}. Other works have motivated an increase in carrier frequencies further into the millimeter-wave region utilizing \SI{150}{\giga\hertz} and \SI{300}{\giga\hertz} carriers and demonstrating sub-degree angular resolution in forward-looking automotive \ac{sar} applications utilizing a rotating scanning beam \cite{gishkori2018imaging, gishkori2019imaging}. More recently, the possibility of using cooperative \ac{sar} at millimeter-wave has been further suggested to increase look angle diversity, and thus angular resolution in an \ac{its} application \cite{tagliaferri2021cooperative}.} Recently, interest in utilizing polarimetry\hlb{, the measurement of the polarization of scattered waves in a scene,} in automotive radar has also increased based on the desire to obtain more scene awareness for various automated driving functions such as landmark classification and object identification \cite{visentin2017polarimetric, weishaupt2019polarimetric}, ego-vehicle localization \cite{weishaupt2020polarimetric}, multi-path detection and close object separation \cite{visentin2017polarimetric}, and scene characterization using polarimetric \ac{sar} in highway driving \cite{jung2020novel}.


In this work, we demonstrate a fully polarimetric \textit{C}-band (\SI{5.9}{\giga\hertz}) automotive radar utilizing a synthetic aperture to reduce the requirement of a single physically large aperture while also utilizing common low frequency \ac{v2x} communications spectrum. \hlb{The system is fully polarimetric, or quad-pol, meaning it can simultaneously measure both co- and cross-polarized scattering from the scene from either transmit polarization, horizontal (H-pol) or vertical (V-pol).  The advantage of a fully polarimetric \ac{sar} system in an automotive application is the ability to infer object types and materials in the scene based on how they affect the polarization of the scattered waves, similarly to the techniques described in \cite{visentin2017polarimetric, weishaupt2019polarimetric, jung2020novel}.} To our knowledge, no prior works have demonstrated \hlb{ground-based} polarimetric \ac{sar} in the \hlb{\textit{C}-band \ac{v2x} spectrum}. This technique leverages the improved angular resolution afforded by synthetic aperture techniques, while utilizing a frequency spectrum previously avoided due to poor spatial resolution concerns. 
The contributions of this work are as follows:
\begin{itemize}
\item[1)] We demonstrate the efficacy of synthetic aperture radar applied to automotive vehicles at a low frequency to avoid the common pitfall of poor angular resolution in low frequency radars, motivating future works on the reuse of existing communication signals for ranging.
\item[2)] We demonstrate the ability of low frequency polarimetric \ac{sar} systems to infer meaningful characteristics about scene geometries encountered in urban driving scenarios.
\end{itemize}
The remainder of the paper is organized as follows: Section \ref{image-formation} describes the basic process behind the \ac{sar} image formation technique used in this paper, Section \ref{experimental-configuration} describes the hardware and processing parameters used to produce the \ac{sar} images, and Section \ref{experimental-results} presents the measured results obtained from each of the scenes, focusing on the unique information obtained at the low frequencies used and  the polarimetric scattering information received.

\section{Synthetic Aperture Radar Image Formation}
\label{image-formation}

\begin{figure}
	\centering
	\subfloat[]{\includegraphics[]{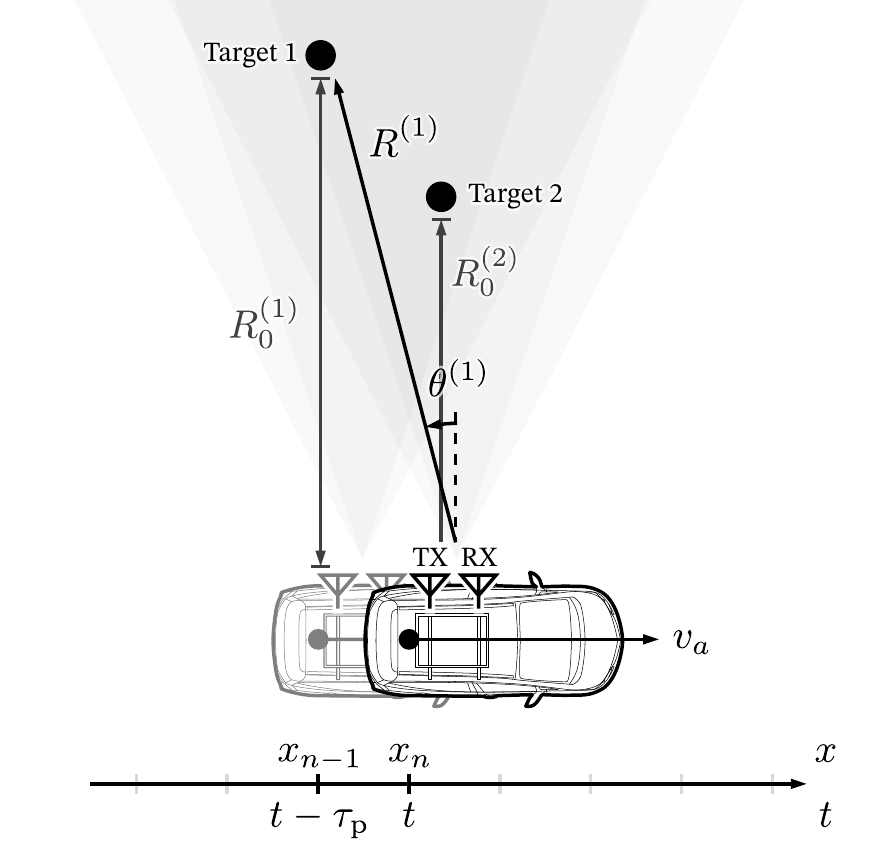}}\\
	\subfloat[]{\includegraphics[]{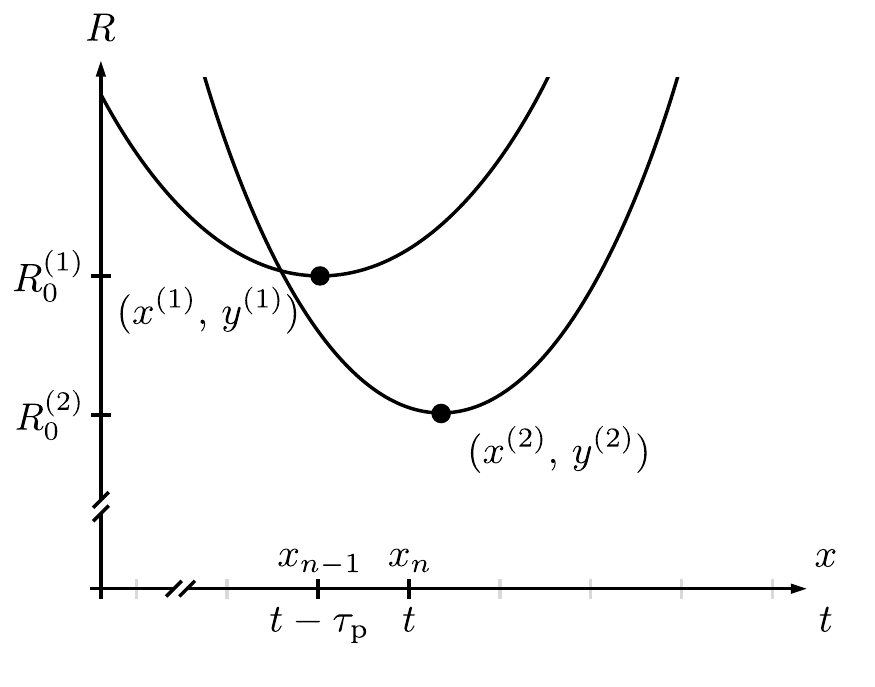}}
	\caption{\hlb{(a) Vehicle and scene geometry for a simplified scene with two point scatterers. Four antenna cones are shown representing the half-power beamwidth of the TX and RX antennas at locations $x_n$ and $x_{n-1}$; differing size antenna beamwidths were used for TX and RX as discussed in Section \ref{experimental-configuration}. A quasi-monostatic approximation is used meaning the TX and RX beams are assumed to be colocated as shown in the figure. (b) Migration of measured target range over time as the platform moves along track traces out a parabola; when the targets reach the \acf{rca} (where $R^{(i)}=R^{(i)}_0$), the downrange ($y$-component) and crossrange ($x$-component) locations are measured at their true locations.}}
	\label{range}
\end{figure}

The primary objective of a \ac{sar} image formation algorithm is to generate a focused image using range measurements taken from multiple locations along a ``synthetic aperture'' track.  Typically, \ac{sar} systems have moderately wide antenna patterns which will collect range measurements from a wide arc of targets in each sample, hence the need for synthesizing a larger aperture.  As the platform moves across the synthetic aperture, the range to the targets in the scene will vary, moving to a \ac{rca} $R_0$ for each target, then moving away, tracing out a parabolic pattern (assuming a linear trajectory) as shown in Fig. \ref{range}.  
This migration in target range over the synthetic aperture length creates a spreading of the targets' energy over range and azimuth bins based on the downrange and crossrange location of the target relative to the \ac{sar} platform track, and thus, produces a location-dependent response to each scatterer in the scene called the \acf{psr}.  The goal of any \ac{sar} image formation system is to focus this energy back to a single point at the proper range and azimuth location while minimizing energy outside of the actual range-azimuth bin at which the target resides.

While there are many algorithms used in \ac{sar} image formation such as \acf{rda}, \ac{rma} (also known as Omega-$K$), \acf{pfa}, and \acf{bpa}, to name a few, \cite{bamler1992comparison, melvin2012principles, jansing2021introduction}, the \ac{rma} \cite{cafforio1991sar, charvat2014small, jansing2021introduction, tolman2008detailed} was chosen for use in our application due to its ability to produce well-focused images over long synthetic apertures
while maintaining a computationally efficient implementation by operating on uniformly spaced samples through the use of a frequency domain Stolt interpolation step. Because of this uniformly spaced data, the image may be formed using a two-dimensional \acf{ifft}. The specific implementation of the \ac{rma} algorithm used in these experiments is based on the approaches used in \cite[Chapter 4]{charvat2014small} and \cite{guo2016modified} which are modified approaches considering side-looking, ground-based applications. There are four major steps to \ac{sar} image formation using the \ac{rma} algorithm: crossrange Fourier transform, downrange matched filtering, Stolt interpolation, and two-dimensional \ac{ifft}. These steps will be briefly discussed in the remainder of this section.


We first discuss the assumptions on platform motion, namely that (1) the target is moving at a constant velocity $v$, and (2) the target is moving linearly. Based on these assumptions, a model for the platform's position during the $n$th pulse can be described using the stop-and-go approximation \cite{meta2007signal, munoz2014beyond} 
as 
\mbox{$x_n=nv\tau_\mathrm{pd}$} where $\tau_\mathrm{pd}$ is the pulse duration, meaning the platform can be modeled as quasi-static during the time it takes for the signal to propagate to the target and back.
This approximation is valid when the total distance the platform moves during a pulse is a small fraction of the carrier wavelength. 
The range of the $i$th target with respect to the platform position can then be expressed as 
\begin{align}
	\label{eq:R}
	R^{(i)}(x_n) = \sqrt{\left(y_\mathrm{t}^{(i)}\right)^2 + \left(x^{(i)}_t-x_n\right)^2}
\end{align}
where $(x^{(i)}_t,\,y^{(i)}_t)$ is the crossrange and downrange position of the $i$th target in the fixed, world reference frame.

A sawtooth, linear \acf{fmcw} ranging waveform was chosen due to the improved range resolution and increased processing gain it enables compared to conventional pulsed waveforms \cite{richards2014fundamentals}, and the ease of implementation in the hardware used for this experiment.
One period of the transmitted 
\ac{fmcw} waveform is simply a linear frequency modulated (\acs{lfm}) waveform and can be represented as
\begin{align}
	s_\mathrm{t}(t) = \alpha\exp{\left\{j2\pi \left(f_0t+\frac{K}{2}t^2\right) \right\}}
\end{align}
where $\alpha$ is a complex amplitude, $f_0$ is the carrier frequency, \mbox{$t\in[-\tau_\mathrm{pd}/2,\,\tau_\mathrm{pd}/2]$}, and \mbox{$K=\beta/\tau_\mathrm{pd}$} where $\beta$ is the \ac{lfm} bandwidth. The received signal is a summation of time-delayed copies of the original signal from all scatterers in the scene, given by
\begin{multline}
	s_\mathrm{r}(x_n,\,t) = \sum_i \alpha_i \\ \exp{\left\{j2\pi \left[ f_0\left(t-\tau_\mathrm{d}^{(i)}(x_n)\right)+\frac{K}{2}\left(t-\tau_\mathrm{d}^{(i)}(x_n)\right)^2\right]\right\}}
\end{multline} 
where $\tau_\mathrm{d}^{(i)}(x_n)=2R^{(i)}(x_n)/c$ is the round-trip time delay of the signal reflected and received from target $i$, and $c$ is the speed of light in the medium. The complex amplitude in the received signal includes the effects of propagation, system losses, and target scattering parameters. Since the principal processing impacts are phase-based, we omit the amplitude terms going forward for brevity. Furthermore, the derivation will be shown for a scene containing a singe point scatterer to elucidate the processing. 

The received signal is dechirped \hlb{by mixing the transmitted signal with the time-delayed receive signal}, yielding a baseband signal that can be easily sampled. To accomplish the dechirping, a copy of the transmitted signal is correlated with the received signal, producing a beat frequency for each received pulse that is proportional to the range of the scatterer. In practice, this process is often physically realized using a hardware mixer\hlb{, which is the technique used in the experiments in this paper}. The downconverted and dechirped signal from a single scatterer is thus represented as 
\begin{multline}
\label{s_rd} 
	s_\mathrm{rd}(x_n, t) = s_\mathrm{r}(x_n,\,t) \cdot s_\mathrm{t}^*(t) = \\ 
	\exp{\left\{ -j\frac{4\pi}{c}(f_0+Kt)R(x_n) \right\}} \exp{\left\{ jK\frac{4\pi}{c^2}R^2(x_n) \right\}}
\end{multline}
where the last term is the unwanted \ac{rvp} due to the dechirping process and represents a static phase that can be easily compensated by multiplying with its conjugate in the Fourier domain; however for the short range nature of automotive \ac{sar} systems, this term may also be neglected with minimal impact. \hlb{It should be noted that a downrange windowing function along the $t$-axis may also be applied here to fit the desired application requirements for mainlobe width and peak sidelobe level \cite[Chapter 14.4]{richards2010principles},\cite[Chapter 7.8]{lathi2005linear}.}


The range term can be expanded \hlb{by substituting \eqref{eq:R}} and, if the stop-and-go platform assumption is used, the compensated baseband signal is represented as
\begin{multline}
	\label{srd_tx}
	s_\mathrm{rd}(x_n,\,t) = \\ \exp{\left\{ -j\frac{4\pi}{c}(f_0+Kt) \sqrt{y_\mathrm{t}^2+\left(x_\mathrm{t}-x_n\right)^2} \right\}}.
\end{multline}
The downconverted signal in \eqref{srd_tx} can then be further simplified by using the round-trip wavenumber representations for the carrier and \ac{lfm} sweep \mbox{$k_r=\omega(t)/c = \frac{4\pi}{c}\left(f_0 + Kt\right)$} where \mbox{$k_r\in\frac{4\pi}{c}[f_0-\beta/2,\,f_0+\beta/2]$}. Thus, after substitutions, the baseband signal is represented as
\begin{align}
	s_\mathrm{rd}(k_r,\,x_n) = \exp{\left\{ -j k_r \sqrt{y_\mathrm{t}^2+\left(x_\mathrm{t}-x_n\right)^2} \right\}}.
\end{align}

\hlb{While the range samples are currently in the spatial frequency domain, the crossrange samples are still a function of time $x_n$. To properly focus the image, Stolt interpolation is performed which requires that the entire image be in the spatial frequency domain. Thus, a} Fourier transform is then applied along the crossrange dimension $x_n$ to obtain the spatial representation of the scene in wavenumber space.
While the transform integral cannot be solved directly in a closed form, the principle of stationary phase is commonly employed to approximate it \hlb{\cite[Chapter 6]{tolman2008detailed}}, yielding
\begin{align}
S(k_r,\,\,k_x) \approx \exp{\left\{ -jy_\mathrm{t}\sqrt{k_r^2-k_x^2} - k_xx_\mathrm{t} \right\}}.
\end{align}

A matched filter is then applied along the range dimension to shift the data to 
a reference range, 
 commonly chosen to be either \mbox{$R_\mathrm{ref}=0$}, or \mbox{$R_\mathrm{ref}=R_\mathrm{max}/2$} for ground-based platforms \cite[Chapter 4]{charvat2014small}, \cite{guo2016modified}. 
The form of the matched filter is
\begin{align}
S_\mathrm{mf}(k_r,\,k_x)=\exp{ \left\{jR_\mathrm{ref}\sqrt{k_r^2-k_x^2} \right\}}
\end{align}
thus the data are shifted in range by $-R_\mathrm{ref}$ leading to
\begin{align}
S(k_r,\,k_x) = \exp{\left\{ j\left[(y_\mathrm{t}-R_\mathrm{ref})\sqrt{k_r^2-k_x^2} - k_xx_\mathrm{t}\right]\right\}}.
\end{align}
Finally, Stolt interpolation \cite{cafforio1991sar, wang2008digital} is applied to linearly map the spatial frequency and wavenumber domains to the spatial wavenumber domain \mbox{$\sqrt{k_r^2-k_x^2}\rightarrow k_y$}, \hlb{in effect, focusing the targets not located at $R_\textrm{ref}$ (as the targets at $R_\textrm{ref}$ are already focused)}, producing
\begin{align}
S(k_y,\,k_x) = \exp{\left\{ j\left[(y_\mathrm{t}-R_\mathrm{ref})k_y - k_xx_\mathrm{t}\right]\right\}}.
\end{align}
After the interpolation, a uniform grid in the $(k_x,\,k_y)$ wavenumber space is formed allowing it to be transformed to a crossrange and downrange representation by a two-dimensional \ac{ift}. After the inverse Fourier transformation, the form of the point responses in the scene can be approximated by sinc functions
\begin{multline}
s(t,\,x_n) \approx \\ \text{sinc}\left\{ \frac{ct-2(y_\mathrm{t}-R_\mathrm{ref})}{\rho_\mathrm{R}} \right\}\text{sinc}\left\{\frac{2(x_n-x_\mathrm{t})}{\rho_\mathrm{CR}}\right\}
\end{multline}
where $\rho_\mathrm{R}$ is the downrange resolution in meters, given by
\begin{align}
	\label{dr_resolution}
	\rho_\mathrm{R}=\frac{c}{2\beta}	
\end{align}
and the crossrange resolution is given by
\begin{align}
	\label{cr_resolution}
		\rho_\mathrm{CR}=\frac{\lambda}{2\theta_\mathrm{3dB}}
\end{align}	
where $\theta_\mathrm{3dB}$ is the $3$-dB beamwidth in azimuth of the antenna \cite{richards2014fundamentals}. 

\hlb{Another important parameter to consider in proper image formation is the sampling criterion: in order to avoid aliasing in the spatial domain, the the phase of the scattered signals must be sampled at least twice per cycle.  To satisfy this, the maximum rate of change of the phase can be found from the Doppler shift of an obstacle passing broadside to the vehicle}
\begin{align}
	\hlb{f_d = \frac{2v_a\sin{\theta}}{\lambda}}
\end{align}	
\hlb{where $f_d$ is the Doppler shift frequency, $v_a$ is the platform velocity, and $\theta$ is the angle of the target off broadside.  The sampling criterion will be based on the maximum angle $\theta$ observed by the system, typically chosen to be $\theta_\mathrm{3dB}/2$ or, more conservatively, $\theta_R$, the Rayleigh (peak-to-null) beamwidth \cite{richards2010principles}.  From this, the following sampling relations are found}
\begin{align}
	\hlb{f_p} &\hlb{> \frac{4v_a \sin{\theta_R}}{\lambda}}\\
	\hlb{\delta_x} &\hlb{< \frac{\lambda}{4 \sin{\theta_R}}}
\end{align}	
\hlb{where $f_p$ is the \ac{prf} and $\delta_x$ is the spatial sampling interval.}

\section{Experimental Configuration}
\label{experimental-configuration}

The experiments conducted in this paper were designed to evaluate both the efficacy of utilizing automotive \ac{v2x} \textit{C}-band communications frequencies, as well as the potential benefits of utilizing polarimetry for inferring scene and object characteristics in urban scenarios.  To accomplish these objectives, an \ac{av} test platform was outfitted with a custom, low-cost, \SI{5.9}{\giga\hertz} \ac{fmcw} radar and polarimetric transmit and receive arrays; several scenes were imaged at Michigan State University's Spartan Village \ac{av} research facility to demonstrate the capabilities of the radar.

\subsection{Radar Hardware}

\begin{figure}[t!]
	\centering
	\includegraphics{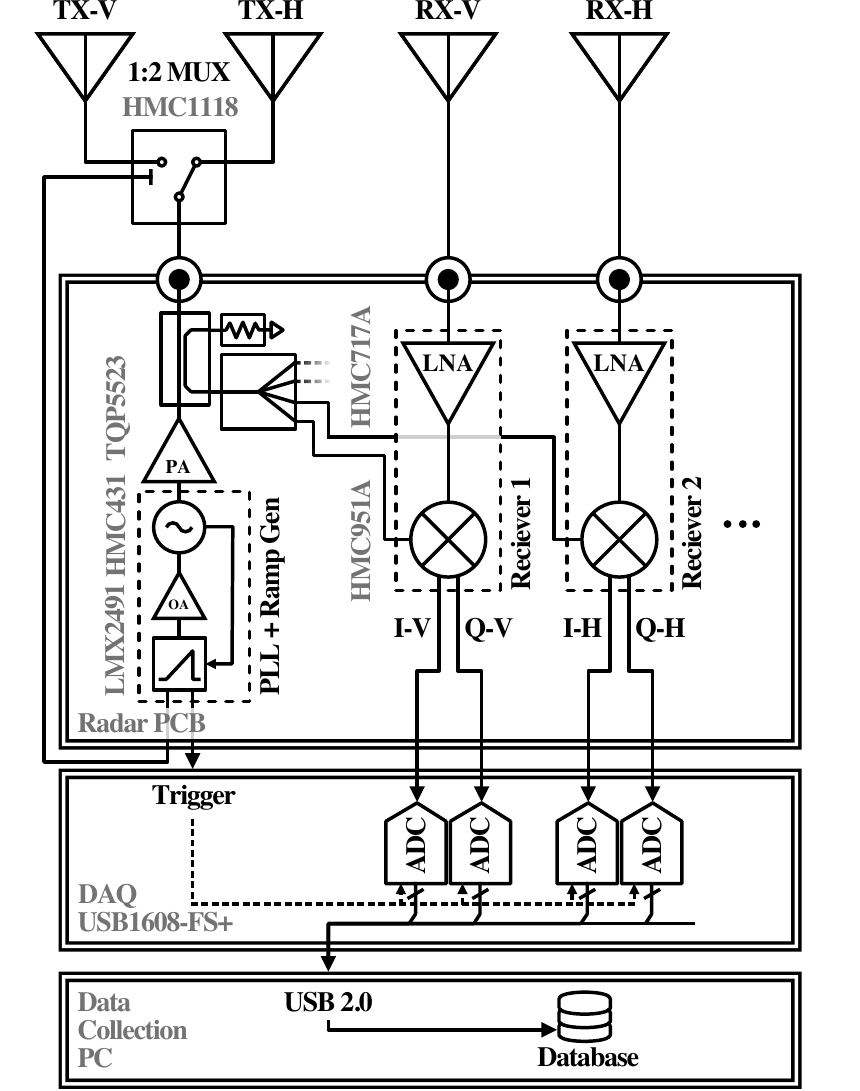}
	\caption{Schematic of the \ac{fmcw} radar \hlb{system} used in these experiments. \hlb{The radar \ac{pcb} (top) was connected to a USB DAQ (middle) to digitize the baseband signals which were then sent to a data collection PC for datalogging and offline processing (bottom).} Part numbers for major components on the \ac{pcb} are shown in grey.  The \ac{pcb} contained one transmit channel and four receive channels, however only two receive channels were used for these experiments. The single transmit channel was multiplexed using an off-board 1:2 switch\hlb{, shown at the top of the figure, to select between horizontal and vertical transmission polarity.}}
	\label{schematic}
\end{figure}

\begin{figure}[t!]
	\centering
	\includegraphics[width=3.44in]{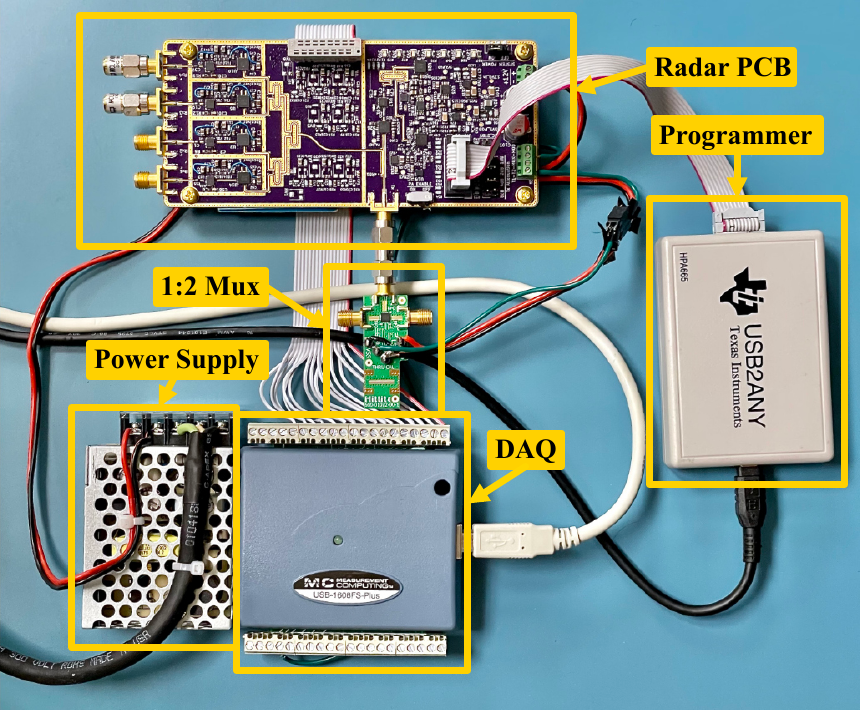}
	\caption{Radar and DAQ configuration used during the experiment \hlb{pictured in a laboratory environment}. The radar \ac{pcb} is shown in the top connected to a 1:2 RF switch used to select the transmit antenna polarization. USB cables from the programmer (right) and DAQ (bottom) were connected to a control laptop in the vehicle. Power was supplied through an in-car AC inverter to AC/DC power supply shown in the lower left of the figure.}
	\label{hardware}
\end{figure}

The radar hardware used for this \ac{sar} imaging experiment consisted of \hlb{an \ac{fmcw} radar \ac{pcb} with a single transmitter and four receivers} in conjunction with a 1:2 multiplexer on the transmitter to obtain time domain multiplexing for the H-pol and V-pol transmissions; \hlb{two of the four receivers were used to receive the V-pol and H-pol scatters simultaneously while the two unused receivers were terminated with matched loads}.  A schematic of the radar \hlb{system} is shown in Fig. \ref{schematic}, and a picture of the full \hlb{radar system} used in the vehicle is shown in Fig. \ref{hardware}.
\hlb{The radar \ac{pcb} was designed in-house as a low-cost open-source radar; design files, software, and build and operation instructions are available in \cite{aps_radar}.} The architecture is direct-downconversion, consisting of a single transmitter and four receivers, and is fabricated on a 4-layer ISOLA FR408-HR substrate (\mbox{$\epsilon_r=3.64,$}\,\mbox{$\tan{\delta}=0.0098$} at $\SI{5.9}{\giga\hertz}$).  The transmitter design centered around the \ac{ti} LMX2491 \SI{6.4}{\giga\hertz} programmable \ac{pll} and ramp generator \ac{ic}. The output of the \ac{pll} was passed through an external active loop filter before entering an \ac{adi} HMC431 \SI{5.5}{\giga\hertz}-\SI{6.1}{\giga\hertz} \ac{vco}. The output of the \ac{vco} was split by a balanced \mbox{$3$-dB} Wilkinson divider where one half of the signal was sent back to the \ac{pll} for reference, and the other half was sent to a variable attenuator before entering a Qorvo TQP5523 \mbox{$32$-dB} \ac{pa}. After the amplifier, the signal passed through a \mbox{$14$-dB} coupler where a small reference signal was split from the amplified signal before leaving the \ac{pcb}. The reference signal then entered a 1:4 network of \mbox{$3$-dB} Wilkinson dividers which were distributed to the receive channels.  Each receive channel consisted of an \ac{adi} HMC717A \mbox{$14.5$-dB} \ac{lna} followed by an \ac{adi} HMC951A \SI{5.6}{\giga\hertz}-\SI{8.6}{\giga\hertz} quadrature downconverter. To accomplish multiplexing of the transmitted signal, an \ac{adi} HMC1118 \SI{100}{\mega\hertz}-\SI{13}{\giga\hertz} SPDT switch evaluation module \hlb{(EVAL-HMC1118)} was connected to the transmitter port on the \ac{pcb}.  The switch was controlled via a ramp start signal on the LMX2491 \ac{pll} ramp generator which would alternate between the H-pol and V-pol for each sequential chirp. Finally, the \ac{if} signals were sampled and digitized using a Measurement Computing USB-1608-FS-Plus \acs{daq} at 100\,kSps/ch which was also triggered by the ramp start signal from the LMX2491.

\subsection{Vehicle Configuration and Data Collection}
\begin{figure}
	\centering
	\includegraphics[width=3.44in]{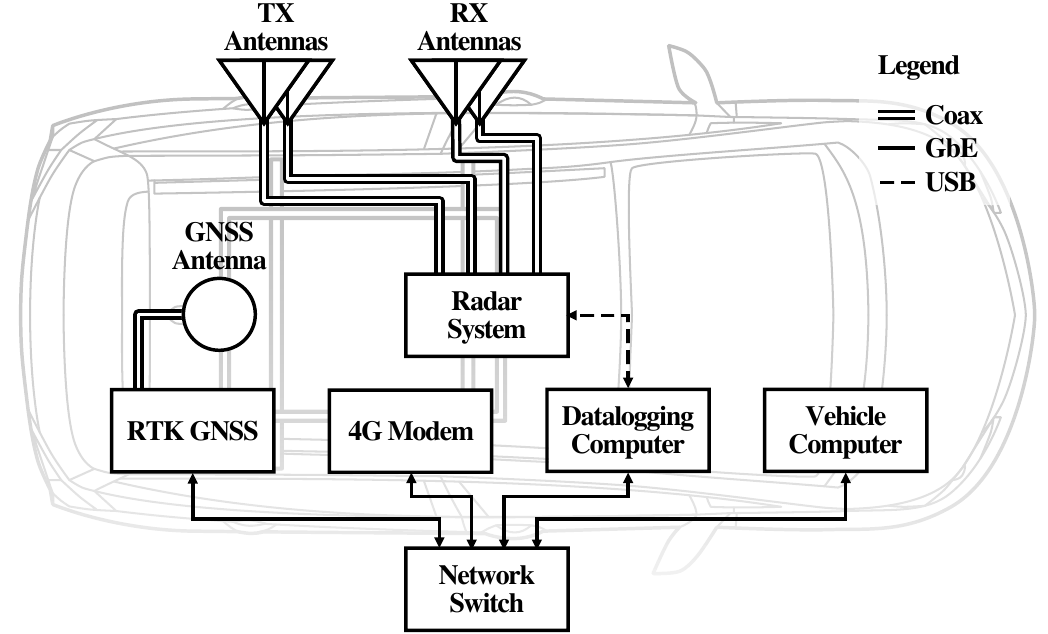}
	\caption{\hlb{Simplified vehicle system architecture detailing hardware used during the \ac{sar} experiments.}}
	\label{vehicle_schematic}
\end{figure}
\begin{figure}
	\centering
	\includegraphics[width=3.44in]{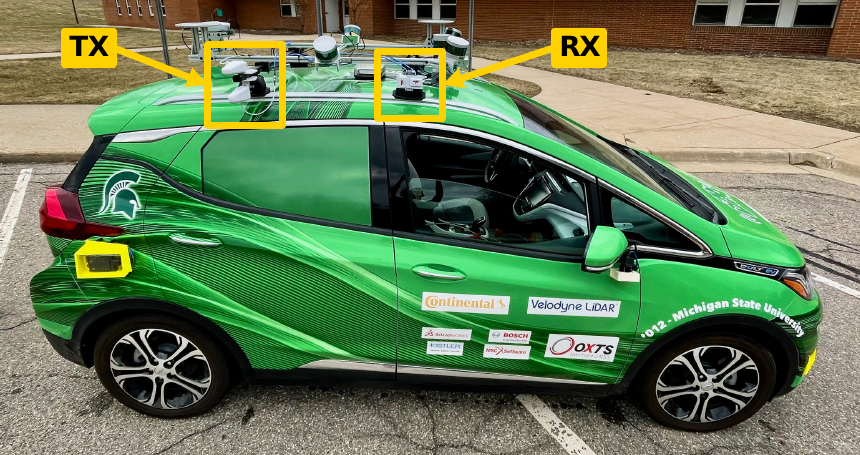}
	\caption{Michigan State University CANVAS Chevrolet Bolt \ac{av} test platform with \ac{sar} antennas located on the roof rack of the vehicle.  The two $13.2$\,dBi Yagi-Uda antennas used for transmit were located towards the rear of the roof rack shown in the upper left of this image; the H-pol antenna is mounted above the V-pol antenna.  The two $9.5$\,dBi Bi-Quad antennas used for receive were located on the front of the roof rack in the upper right of this image; the H-pol is rearwards of the V-pol antenna.}
	\label{vehicle}
\end{figure}

The vehicle used for the imaging experiments was the Michigan State University CANVAS Chevrolet Bolt \ac{av} test platform. \hlb{A simplified vehicle system architecture schematic is show in Fig. \ref{vehicle_schematic} and a photograph of the vehicle is shown in Fig. \ref{vehicle}.} The transmitter array consisted of two $13.2$\,dBi \SI{5.725}{\giga\hertz}--\SI{6.0}{\giga\hertz} Laird Communications Yagi-Uda antennas positioned in a vertical array with the horizontally polarized antenna placed above the vertically polarized antenna.  The receive array consisted of two $9.5$\,dBi \SI{5.0}{\giga\hertz}--\SI{6.0}{\giga\hertz} TrueRC Bi-Quad antennas placed in a horizontal linear array where the vertically polarized antenna was placed slightly forward of the horizontally polarized antenna on the roof rack of the vehicle. \hlb{Ideally, both the transmit and receive antennas would be single dual polarized antennas with identical beamwidths, however the antennas were chosen due to availability at the time of the experiment.}  Both the transmit and receive arrays were positioned at approximately the same elevation from the ground.

For each experiment, the vehicle's position was measured by a Swift Navigation Piksi Multi \acf{rtk} \acf{gnss} unit to provide position and velocity information for \ac{sar} image formation. The \ac{gnss} antenna was also located on the roof rack in the center of the vehicle track, near the rear of the vehicle's wheelbase. \hlb{The system received real-time \acf{cors} correction data for \ac{rtk} phase corrections from the \ac{mdot} basestation in Lansing via an onboard 4G modem. The positional information provided was observed to have a typical standard deviation of about $\SI{20}{\milli\meter}$ ($0.4\lambda$ at \SI{5.9}{\giga\hertz}) in the experiment location at the time the data collection occurred.} 

\begin{figure*}
  \centering
  \includegraphics{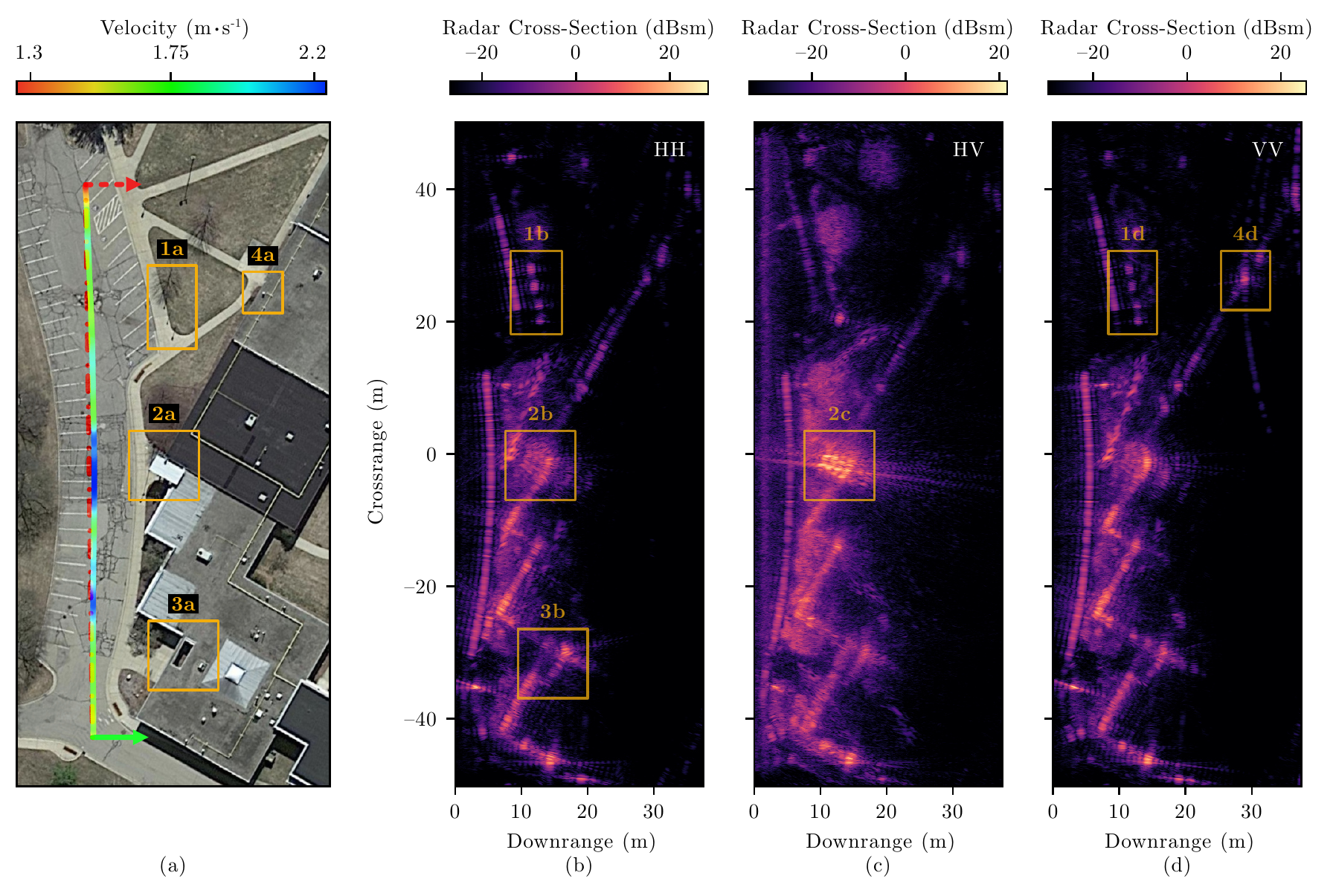}
  \caption{Optical satellite image of \ac{sar} track one (a) with the vehicle's trajectory shown\hlb{; the dot-dashed red line shows the approximated trajectory used for image formation---due to imperfect vehicle control and changes in parking lot contour, the actual path deviates slightly.} The solid green arrow indicates track start, dashed red arrow indicates track end. The arrow's pointing direction indicates the radar's look vector. Images are presented for HH (b), HV (c), and VV (d). Received intensity is calibrated to a 4.41-dBsm corner reflector located at $\approx{\SI{21}{\meter}}$ crossrange and $\approx{\SI{13}{\meter}}$ downrange in the bottom of \ac{roi} 1. The track hue represents the relative velocity throughout the track, red indicates relatively low velocity segments while green to blue represents medium to high velocity. \ac{roi} 1 shows three traffic signs and the corner reflector used for calibration. \ac{roi} 2 highlights the region where a metallic, corrugated awning which produces a large cross-polarized response is located (a ground-based optical image of awning is shown in Fig. \ref{awning}). \ac{roi} 3 highlights an opening in the roof. \ac{roi} 4 shows a vertical metallic building exhaust chimney which produces a large vertically polarized response. Maps Data: Google, Imagery \copyright2021, Maxar Technologies.}
  \label{sar_67} 
\end{figure*}

The processing of the data occurred offline using Python. The trajectories driven were plotted and overlaid using Google Maps satellite imagery and are shown in Figs. \ref{sar_67}a, \ref{sar_71}a, and \ref{sar_74}a.  Across all runs, the average maximum velocity was measured to be an approximately \SI{2.3}{\meter\cdot\second^{-1}} ($5.2\,$mph) with a minimum average velocity of about \SI{1.4}{\meter\cdot\second^{-1}} ($3.1\,$mph).  While the trajectories were intended to be nearly straight, there were subtle deviations from linear which, in addition to changing roll and elevation induced by the uneven driving surfaces, caused some de-focusing effects in the final processed \ac{sar} images.

During the experiments, the radar was configured with an 
\ac{lfm} center frequency of $f_0=\SI{5.9}{\giga\hertz}$, an \ac{lfm} bandwidth of $\beta=\SI{200}{\mega\hertz}$, and an \ac{lfm} pulse duration of $\tau_\mathrm{pd}=\SI{1.0}{\milli\second}$.  The radar was configured to continuously transmit sawtooth up-chirp waveforms in an \ac{fmcw} fashion with a nominal \ac{prf} of $f_\mathrm{p(nom)}=\SI{1}{\kilo\hertz}$, however due to hardware limitations of the \acs{daq}, the data needed to be polled by the host computer in software which limited the effective \ac{prf} to $f_\mathrm{p}\approx{\SI{75}{\hertz}}$\hlb{, or effective \ac{pri} of $\tau_p=\SI{13.3}{\milli\second}$}. To maintain approximate temporal alignment between the V-pol and H-pol transmit chirps, the sequential orthogonal polarity chirps were collected using a burst function on the \acs{daq} to ensure there would be collected sequentially, with no time delay between them.

Based on the specified parameters, the minimum downrange resolution was found to be $\rho_\mathrm{R}=\SI{750}{\milli\meter}$, limited by the radar modulation bandwidth; the minimum theoretical crossrange resolution was found to be $\rho_\mathrm{CR} \approx \SI{37}{\milli\meter}$ which is limited by the beamwidth of the higher gain $13.2$-dBi transmit antennas used.
A full summary of the \ac{sar} hardware and processing parameters are shown in Table \ref{tab:sar-parameters}.

\begin{table}[tb]
	\caption{\Ac{sar} Parameters}
	\label{tab:sar-parameters}
  	\begin{tabularx}{\columnwidth}{p{0.55\linewidth}YY}
	
	\toprule[1pt]
	\textbf{Waveform Parameters} \\
	\midrule
	\midrule
	Parameter & Symbol & Value \\
	\midrule
	Carrier Frequency & $f_0$ & \SI{5.9}{\giga\hertz} \\
	Bandwidth & $\beta$ & \SI{200}{\mega\hertz} \\
	Pulse Duration & $\tau_\mathrm{pd}$ & \SI{1.0}{\milli\second} \\
	\Acf{prf} & $f_\mathrm{p}$ & \SI{75}{\hertz} \\
	\Acf{pri} & $\tau_\mathrm{p}$ & \SI{13.3}{\milli\second} \\
	\midrule[1pt]
	\textbf{Hardware Parameters}\\
	\midrule
	\midrule
	Parameter & Symbol & Value \\
	\midrule
	Approximate Radar Velocity & $v_\mathrm{a}$ & \SI{2.25}{\meter\cdot\second^{-1}}\\
	A/D Sampling Rate & $f_\mathrm{s}$ & $100$\,Ksps \\
	Transmit Power & $P_\mathrm{tx}$ & $+30$\,dBm \\
	\midrule[1pt]
	\textbf{Antenna Parameters}\\
	\midrule
	\midrule
	Parameter & Symbol & Value \\
	\midrule
	Transmit Antenna Gain & $G_\mathrm{tx}$ & $13.2$\,dBi\\
	Receive Antenna Gain & $G_\mathrm{rx}$ & $9.5$\,dBi\\
	Transmit Antenna Half-Power Beamwidth & $\theta_\mathrm{3dB,tx}$ & \ang{40}\\
	Receive Antenna Half-Power Beamwidth & $\theta_\mathrm{3dB,rx}$ & \ang{65}\\
	\midrule[1pt]
	\textbf{Imaging Parameters} \\
	\midrule
	\midrule
	Parameter & Symbol & Value \\
	\midrule
	Maximum Unambiguous Range & $R_\mathrm{max}$ & \SI{37.5}{\meter} \\
	Range Resolution & $\rho_\mathrm{R}$ & \SI{750}{\milli\meter}\\
	Crossrange Resolution & $\rho_\mathrm{CR}$ & \SI{37}{\milli\meter} \\
	\hlb{Spatial Sampling Interval} & \hlb{$\delta_x$} & \hlb{\SI{30}{\milli\meter}} \\
	\bottomrule[1pt]
	\end{tabularx}
\end{table}

\begin{figure}
  \centering
  \includegraphics[width=3.5in]{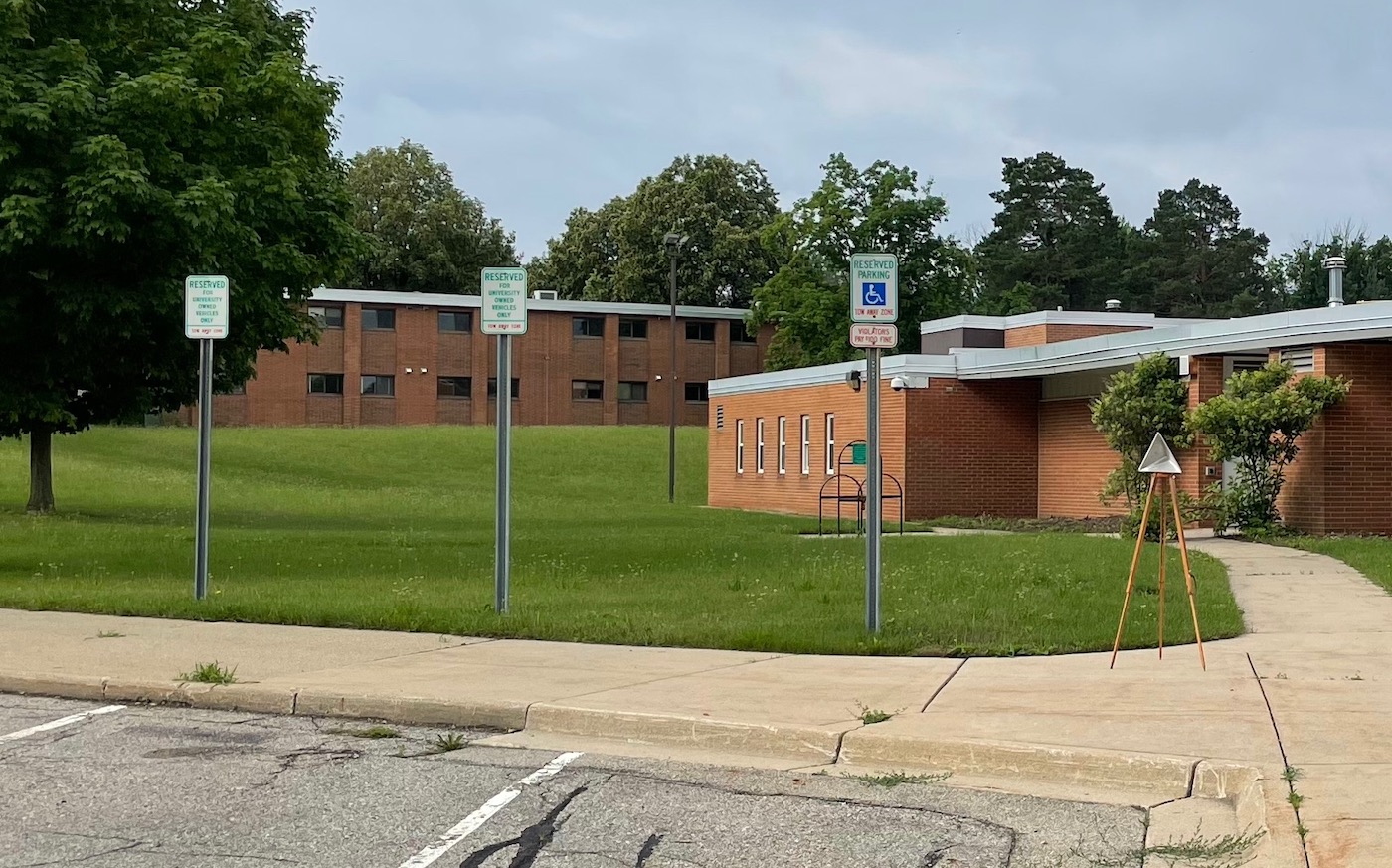}
  \caption{Ground-based image of the row of parking signs and corner reflector used for calibration as well as the metallic exhaust chimney highlighted in \acp{roi} 1 and 4 of Fig. \ref{sar_67}. The cylindrical exhaust chimney is located on the roof of the building in the upper right of the image.}
  \label{signs} 
\end{figure}

\begin{figure}
  \centering
  \includegraphics{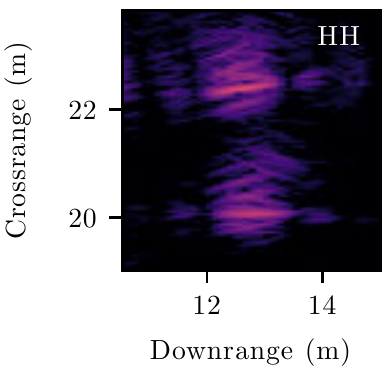}
  \caption{Cropped image of the corner reflector and the first traffic sign from \ac{sar} track one (optical image shown in Fig. \ref{signs}) detailing the focusing and angular resolution of the \ac{sar} image.}
  \label{resolution} 
\end{figure}

\begin{figure}
  \centering
  \includegraphics[width=3.5in]{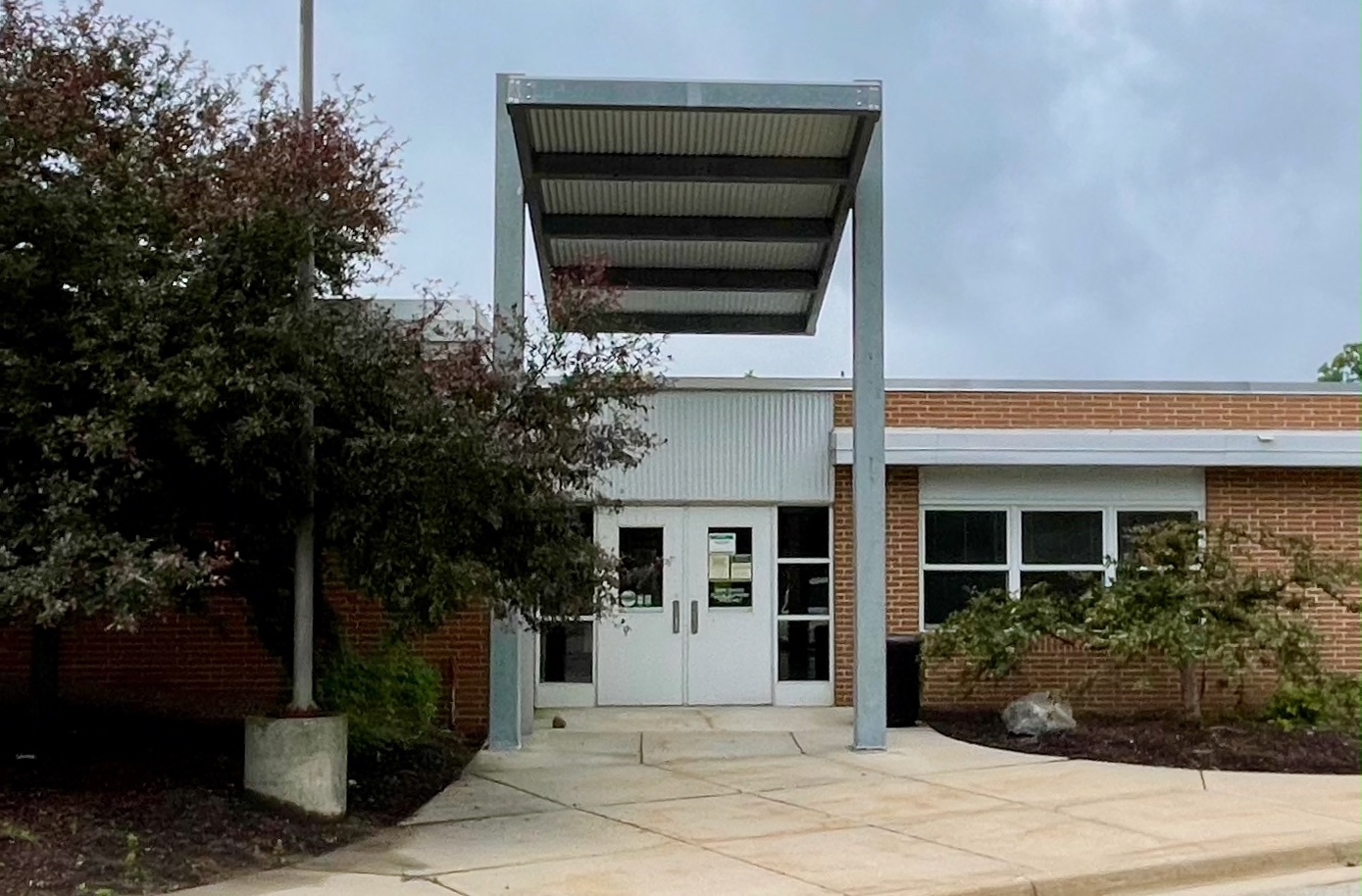}
  \caption{Ground-based image of the awning highlighted in \ac{roi} 2 of Fig. \ref{sar_67}.}
  \label{awning} 
\end{figure}

\section{Experimental Results}
\label{experimental-results}


\begin{figure}
  \centering
  \includegraphics{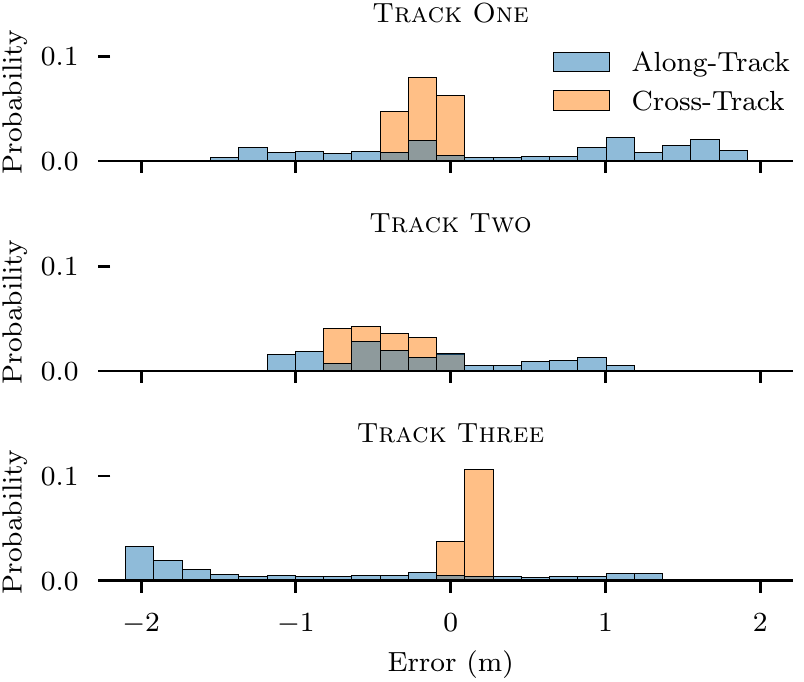}
  \caption{\hlb{Cross-track and along-track error probability distributions for each of the \ac{sar} tracks.}}
  \label{track_error} 
\end{figure}

\begin{figure}
  \centering
  \includegraphics{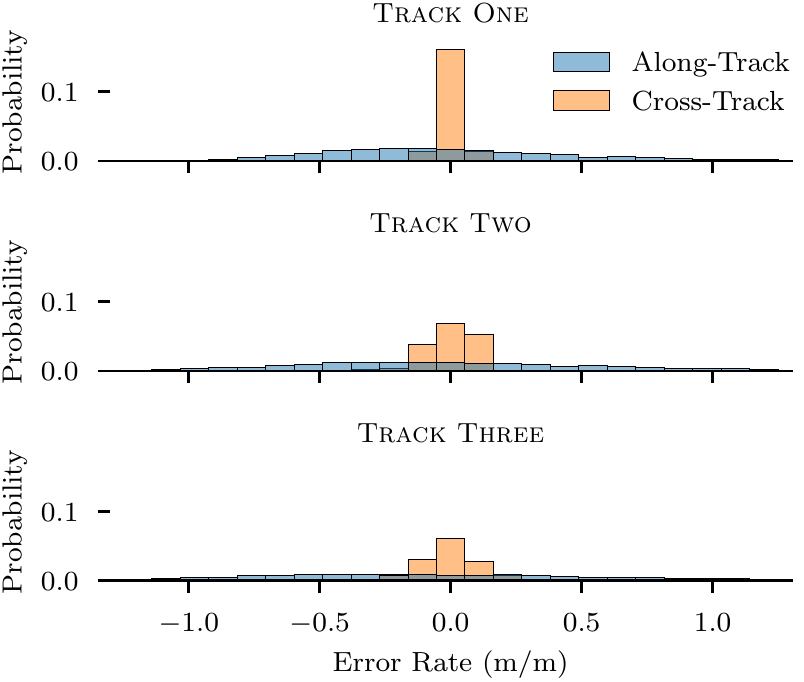}
  \caption{\hlb{Cross-track and along-track error-rate probability distributions for each of the \ac{sar} tracks.}}
  \label{track_error_rate} 
\end{figure}


Three scenes from Michigan State University's Spartan Village \ac{av} testing facility were chosen to highlight unique attributes of the low frequency polarimetric \ac{sar} system. Figs. \ref{sar_67}, \ref{sar_71}, and \ref{sar_74} show optical satellite images of each of the tracks, including the trajectory and vehicle velocity followed during the sample, along side the HH, HV, and VV polarized \ac{sar} images---VH imagery was omitted as the results were found to be very similar to HV for the scenes imaged in these experiments. In each experiment, the positional \ac{rmse} was computed to describe the deviation in position due to cross-tracking errors, and velocity errors due to the constant velocity assumption\hlb{; additionally, along-track and cross-track error and error rate histograms are provided in Figs. \ref{track_error} and \ref{track_error_rate} to provide further insight into the nature of the approximation error in each track where are primarily in the along-track direction. While there exist large-scale errors on the order of meters across the track which result in a large-scale distortion of the image, the errors over small distances are relatively minor, leaving the smaller features intact. This could be corrected for using motion compensation techniques discussed later in Section \ref{discussion}.} The \ac{sar} images for each of the three tracks shown were generated using the \ac{rma} process as described in Section \ref{image-formation} where, for each of the polarizations, the notation is transmit polarization followed by receive polarization, i.e. for HV, a horizontally polarized antenna was used to transmit the waveform while a vertically polarized antenna was used to receive its scattered energy.  Labeled \acfp{roi} are also included as yellow boxes to highlight features from the scene that are discussed in the text. To estimate the reflected power intensity, a $4.41$-dBsm corner reflector was placed along each track as a reference calibration target.  Finally, it should be noted that the optical satellite imagery included was taken at a time when the leaves were not present on the trees and bushes, however the experiments were conducted with the foliage present.

\subsection{Synthetic Aperture Track One}

\ac{sar} track one is shown in Fig. \ref{sar_67}. In this track the vehicle maintained a speed of between \SI{1.3}{\meter\cdot\second^{-1}} ($2.9$\,mph) and \SI{2.2}{\meter\cdot\second^{-1}} ($4.9$\,mph); using the constant-velocity model, the vehicle obtained \hlb{a positional approximation} \ac{rmse} of \SI{0.87}{\meter}.  In addition to the error induced by heading and velocity variation, the elevation of the parking lot increased and sloped right-to-left along the track which were not compensated for in these images, possibly contributing to further focusing errors. The vehicle's trajectory and optical scene image can be seen in Fig. \ref{sar_67}a.

The \ac{sar} images for track one are shown in Figs. \ref{sar_67}b--\ref{sar_67}d.   In all three images, the oblique outline of the building and parking lot curb can be clearly seen; even smaller details are visible along the building such as the small skylight aperture in the building roof shown in \ac{roi} 3.  Another interesting area to note is \ac{roi} 1b and 1d in the HH and VV polarization images.  These sections illustrate a row of three parking signs positioned next to the corner reflector used for image calibration and provide a good location for investigating the image focusing performance. An optical ground-based image of these targets is shown in Fig. \ref{signs}. Each of these objects show up very clearly in the co-polarized images, while only the corner reflector is present in the cross-polarized image. Regarding the realized resolution, a zoomed image displaying the corner reflector and first street sign is shown in Fig. \ref{resolution}; while the main lobe of the corner reflector appears to be on the order of \SI{10}{\centi\meter} across, a large amount of the energy is still spread over a region nearly \SI{0.5}{\meter} across. While the spreading of energy in crossrange is not ideal, techniques for mitigating this defocusing will be discussed \hlb{in Section \ref{discussion}}. Also of note is the cylindrical metal building exhaust chimney labeled in \ac{roi} 4 and shown optically from the ground in Fig. \ref{signs}.  This feature is particularly tall and reflective and thus appears strongly in the VV polarization image.  Another particularly interesting area is \ac{roi} 2, shown optically in a ground-based image in Fig. \ref{awning}; here the cross-polarized responses from HV and VH were very strong---this large cross-polarized is believed to be due to scattering off the corrugated metal structure used in the awning and above the doorway (shown in Fig. \ref{awning}).  Finally, the region imaged just above the awning in the \ac{sar} image (or to the left in the ground-based image in Fig. \ref{awning}) shows dense plant foliage; in the \ac{sar} images the wall behind the foliage is still clearly visible, which illustrates the unique ability for the lower frequency waves to propagate through dense foliage like bushes, which are commonly used in urban landscaping. At the higher \SI{76}{\giga\hertz} frequency, most of the power is reflected off of foliage leaving broad shadows behind bushes near the ground \cite{feger2017experimental}.  It should be noted that the ground rises significantly underneath the tree as well which is believed to cause the bulk scattering seen in this area in the VV and HH images. \hlb{Another area to note is the relatively high sidelobe levels on strong scatterers such as building corners, curbs, and metal posts---this is due to the choice of not windowing the sampled data to provide an unbiased view of the ranging performance. As noted in Section \ref{image-formation}, a downrange window may be used to meet application-specific requirements.}

\subsection{Synthetic Aperture Track Two}
\begin{figure*}
  \centering
  \includegraphics{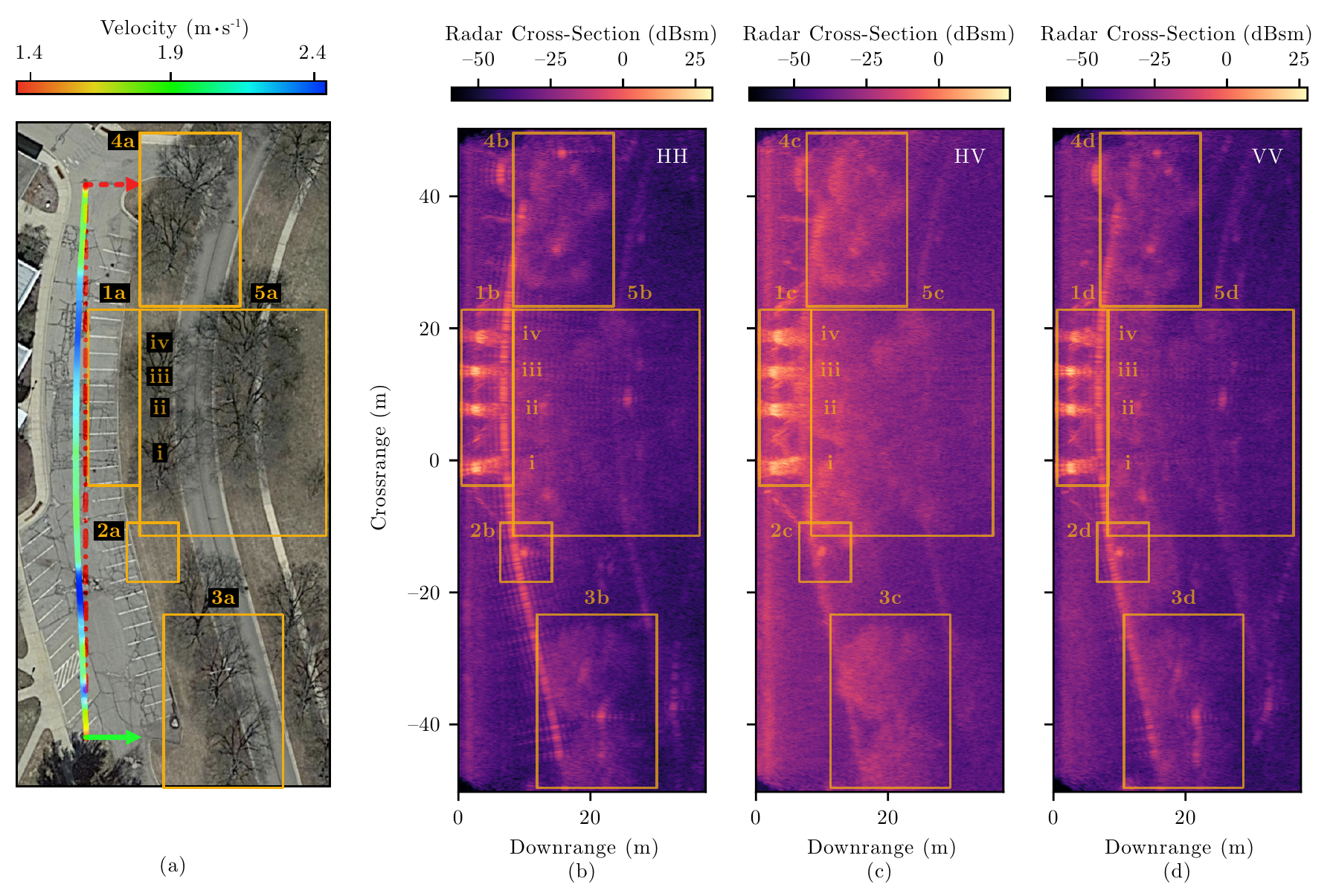}
  \caption{Optical satellite image of \ac{sar} track two (a) with the vehicle's trajectory shown\hlb{; the dot-dashed red line shows the approximated trajectory used for image formation---due to imperfect vehicle control and changes in parking lot contour, the actual path deviates slightly.} The solid green arrow indicates track start, dashed red arrow indicates track end. The arrow's pointing direction indicates the radar's look vector.  Images are presented for HH (b), HV (c), and VV (d). Received intensity is calibrated to a 4.41-dBsm corner reflector located \ac{roi} 2. The track hue represents the relative velocity of the vehicle throughout the track, red indicates relatively low velocity segments while green to blue represents medium to high velocity. \ac{roi} 1 shows a row of four vehicles parked in the parking spaces\hlb{; vehicles are numbered i--iv and correspond with the measurements shown in Table \ref{tab:vehicle-measurements}}. \ac{roi} 3 shows two trees with a metallic street light in between them. \ac{roi} 4 shows two trees with a stop sign above them. \ac{roi} 5 shows four trees, two on the near side of the street, two on the far side, with a street light between the trees on the far side. Maps Data: Google, Imagery \copyright2021, Maxar Technologies.}
  \label{sar_71} 
\end{figure*}

\begin{figure*}
  \centering
  \includegraphics[width=\textwidth]{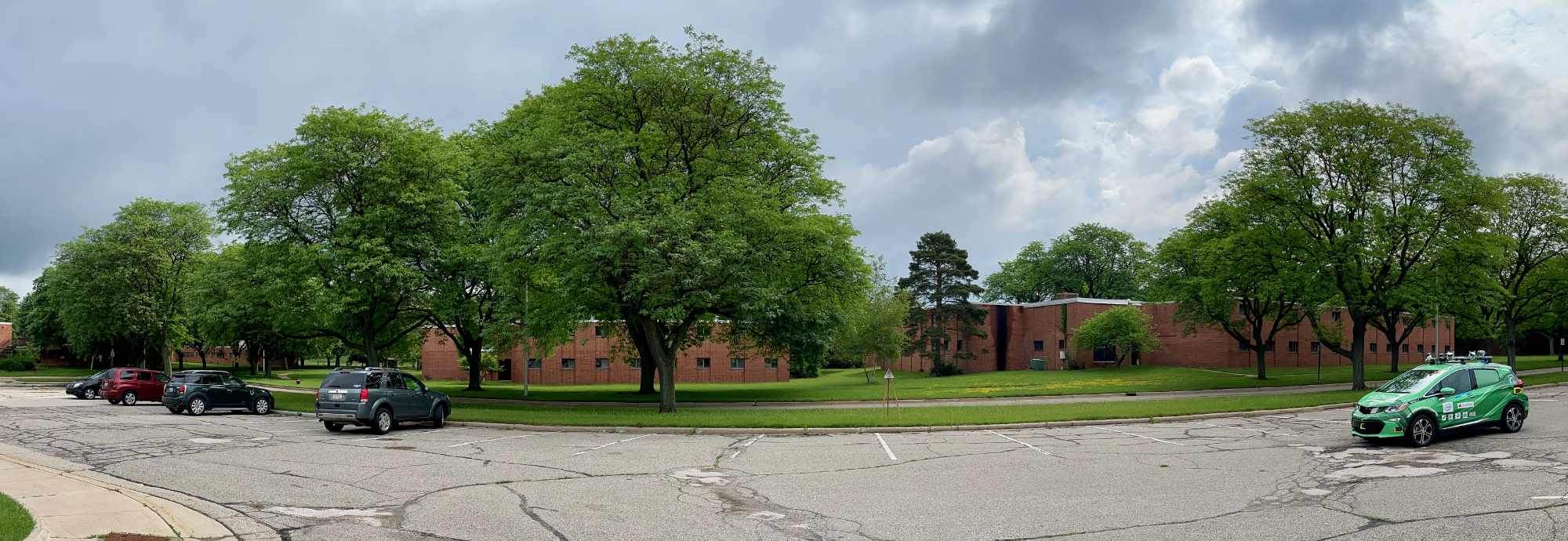}
  \caption{Ground-based optical image of \ac{sar} image track two. The corner reflector is shown in the center of the image. A row of parked cars is shown on the left; the first and second cars from the right have two empty spaces between them while the rest of the vehicles have only a single empty space separating them.}
  \label{sar_71_image} 
\end{figure*}

\begin{table}[tb]
	\caption{\hlb{Vehicle Measurements}}
	\label{tab:vehicle-measurements}
  	\begin{tabularx}{\columnwidth}{YYY}
	
	\toprule[1pt]
	Vehicle Number & \multirow{2}{*}{Width Excl. Mirrors (m)} & \multirow{2}{*}{Measured Width (m)} \\
	(See Fig. \ref{sar_71}) \\
	\midrule
	iv & $1.78$ & $1.74$ \\
	iii & $1.75$ & $1.54$  \\
	ii & $1.83$ & $1.68$  \\
	i & $1.83$ & $1.75$  \\
	\bottomrule[1pt]
	\end{tabularx}
\end{table}

Track two, shown in Fig. \ref{sar_71}, was chosen to show a parking area with vehicles parked in it, a road where the road edge can be seen in the \ac{sar} image, and an illustration of the ability to differentiate trees from man-made structures using polarimetric scene information.  During track two, the vehicle maintained a speed of between \SI{1.4}{\meter\cdot\second^{-1}} ($3.1$\,mph) and \SI{2.4}{\meter\cdot\second^{-1}} ($5.4$\,mph); using the constant-velocity model, a positional \hlb{approximation} \ac{rmse} of \SI{1.01}{\meter} was achieved.  Similarly to \ac{sar} track one, the parking lot had a grade, however in this track, the grade was from left-to-right at the beginning and flattened near the end.  Additionally, a more pronounced curvature in the trajectory driven can be seen in Fig. \ref{sar_71}a, moving from left-to-right along the track shown.  A ground-based optical image of the scene imaged in track two is shown in Fig. \ref{sar_71_image}.

The \ac{sar} images for track two are shown in Figs. \ref{sar_71}b--\ref{sar_71}d. All of the \ac{sar} images show the curb of the parking lot and the far side of the road to some extent, however they are clearest in the HH polarization, as is expected for a predominantly horizontal scattering surface. The most pronounced region of the scene is the region with the four vehicles imaged in the parking spaces in \acp{roi} 1b--1d and shown in the left of Fig. \ref{sar_71_image} and is clearly visible in all polarization images.  The top three vehicles each had a single space between them while the lower vehicle had two spaces between it and the rest of the vehicles.  \hlb{As a figure of measurement accuracy, the measured width of each vehicle was compared with the specified width of the vehicle, excluding mirrors, and is summarized in Table \ref{tab:vehicle-measurements}.  The measured width was found by selecting the approximate crossrange location of the peak downrange intensity at the lateral extrema of the vehicles.  While the measured widths are consistently underestimated, this is not unexpected due to the specular nature of the vehicles at \textit{C}-band wavelengths.  It is anticipated that as the \ac{sar} platform passes the edges of the vehicles, most of the energy is reflect away from the \ac{sar} platform into the environment due to the low grazing angle, or refracted around the edges of the vehicle, resulting in an underestimation of vehicle width.}  The clearly visible vehicle, parking lot, and road geometries demonstrate the ability to easily produce an occupancy map with fine enough resolution to determine parking space vacancy. The corner reflector shown in \ac{roi} 2 can be seen across all polarizations, similar to scene one. \ac{roi} 4 shows two trees situated at the top of the \ac{sar} track with a stop sign located just above them. \ac{roi} 5 shows four trees, two on either side of the road, with a street light post on the far side of the road situated between the two trees. These images highlight the unique ability for the HV polarization image to be used to isolate foliage from man-made objects, like street signs, or other upright objects such as the trunks of the trees. The entire foliage of the trees are shown as more intense HV scatterers illustrating the ability for the waves to propagate through the tree canopy, however the near-side perimeter is the most intense reflection, as is expected. In \ac{roi} 4 the use of the polarization diverse images can be used to infer that there are two trees and another upright object in the \ac{roi} based on the foliage outlines in the HV image, and the three upright scatterers underneath the trees; the tree trunks are also located very near the center of each of the tree foliage outlines in the HV image, making it possible to infer if a scatterer is part of a tree or another object. A similar approach could be used for isolating the light post from the four trees in \ac{roi} 5, however the reflected power from the tree trunks at the far side of the street was much lower, making these trunks more difficult to detect, though still visible to the eye in the \ac{sar} images.

\subsection{Synthetic Aperture Track Three}
\begin{figure*}
  \centering
  \includegraphics{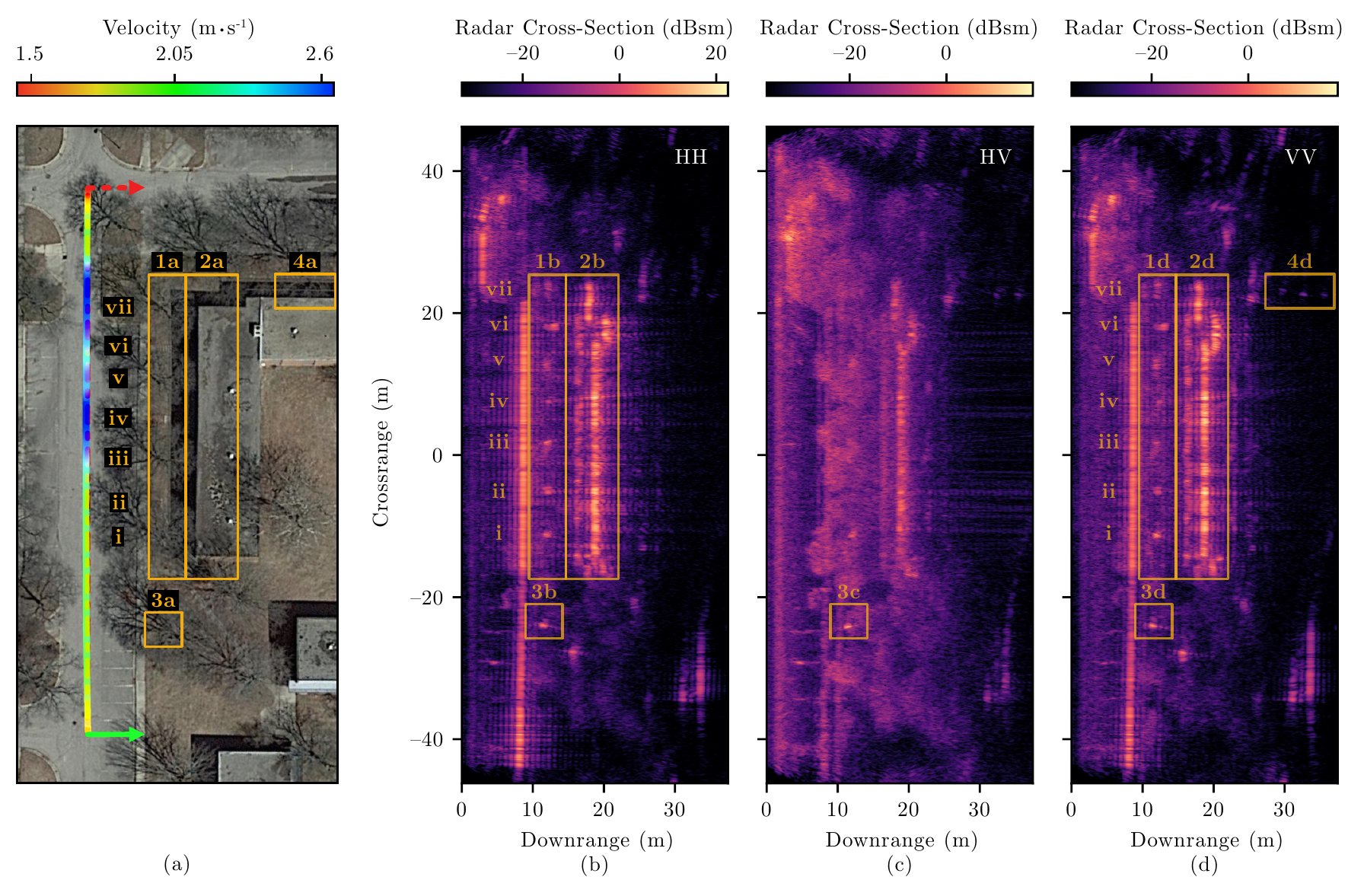}
  \caption{Optical satellite image of \ac{sar} track three (a) with the vehicle's trajectory shown\hlb{; the dot-dashed red line shows the approximated trajectory used for image formation. Due to the highly linear trajectory the vehicle followed on this run, the dot-dashed line is mostly hidden under the colored vehicle trajectory path.} The solid green arrow indicates track start, dashed red arrow indicates track end. The arrow's pointing direction indicates the radar's look vector.  Images are presented for HH (b), HV (c), and VV (d). Received intensity is calibrated to a 4.41-dBsm corner reflector located in \ac{roi} 3. The track hue represents the vehicle's relative velocity throughout the track, red indicates relatively low velocity segments while green to blue represents medium to high velocity. \ac{roi} 1 shows a row of trees and steps embedded in the sidewalk; \acp{roi} 1-i, 1-iii, 1-v, and 1-vii are tree trunks, while the \acp{roi} 1-ii, 1-iv, and 1-vi are steps. \ac{roi} 2 shows two strong linear crosstrack scatterers; the first is due to the poles and fence at the edge of the balcony, and the second is due to the wall of the building. \ac{roi} 4 shows vertical posts on the far side of the building. Maps Data: Google, Imagery \copyright2021, Maxar Technologies.}
  \label{sar_74} 
\end{figure*}

\begin{figure*}
  \centering
  \includegraphics[width=\textwidth]{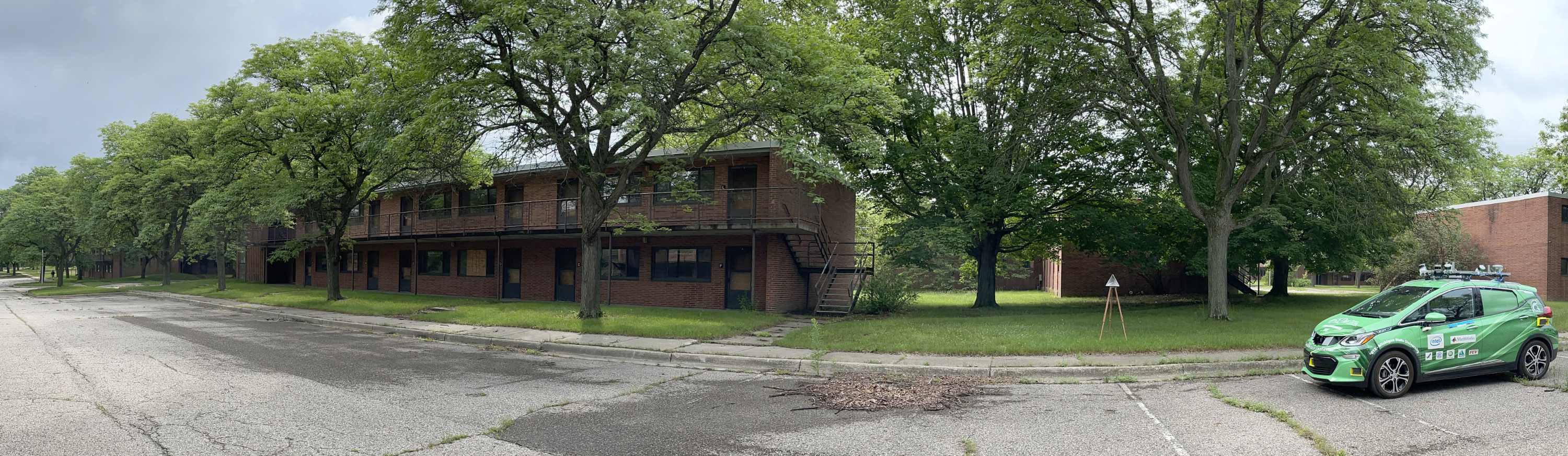}
  \caption{Ground-based optical image of \ac{sar} track three. The corner reflector is shown near the right side of the image. Small steps can be seen in the sidewalk leading up to the building between the trees on the left side of the image.}
  \label{sar_74_image} 
\end{figure*}

\ac{sar} track three is shown in Fig. \ref{sar_74}. During track three, the vehicle maintained a speed of between \SI{1.5}{\meter\cdot\second^{-1}} ($3.4$\,mph) and \SI{2.6}{\meter\cdot\second^{-1}} ($5.8$\,mph).  A slight curve from left to right was exhibited in this track as well, similar to track two.  Using the constant velocity model, a positional \hlb{approximation} \ac{rmse} of \SI{1.01}{\meter} was obtained.  A ground-based optical image of track two is shown in Fig. \ref{sar_74_image}.

The \ac{sar} images for track three are shown in Figs. \ref{sar_74}b--\ref{sar_74}d. The corner reflector used for scene calibration can be seen in \ac{roi} 3 across all polarizations.  \ac{roi} 1 shows an interesting alternating pattern of tree trunks and cement steps embedded in the sidewalk.  To differentiate the scatterers, roman numerals were assigned to each---odd numerals are scatterers from tree trunks while even numerals were associated with the steps embedded in the sidewalk.  These steps can be seen in Fig. \ref{sar_74_image} in the sidewalk leading towards the building in the center-left of the scene; each of these sidewalks had two small steps at a range of $\approx$${\SI{12}{\meter}}$ downrange.  These can be clearly seen in both the co-polarized images, however they are not present in the HV image which primarily shows only foliage with a weak response around the building's wall.  \ac{roi} 2 shows the region of the image including the wall of the building and the railing and posts supporting the balcony.  The \ac{roi} contains two intense regions along the crossrange at downrange distances of $\approx$${\SI{15}{\meter}}$ and $\approx$${\SI{17}{\meter}}$ followed by a third small reagion near the top of the \ac{roi} at a downrange of $\approx$${\SI{20}{\meter}}$ representing an alcove in the building. The nearer of the two intense crossrange regions is due to the near edge of the balcony, the balcony railing, and the posts supporting the balcony.  The eight posts supporting the balcony can be seen clearly on close inspection especially in the VV image. The farther, intense crossrange scatterer is due to the wall of the building. Finally, \ac{roi} 4 illustrates the vertical posts supporting the balcony on the far side of the building, detected in the VV image.

\subsection{Discussion}
\label{discussion}

Based on the images generated from tracks 1--3 it is clear that \ac{sar} is a viable option for generating high resolution imagery from an \ac{av} in the low frequency \ac{v2x} communication spectrum. It is also clear that the long wavelengths afforded by the lower \ac{v2x} bands provide the unique ability to penetrate dense foliage, whereas typical \SI{76}{\giga\hertz} automotive frequencies are fully occluded by vegetation, motivating the use of lower frequencies in automotive radar for detection of hidden hazards.  Furthermore, the use of polarimetry is demonstrated to be able to aide in the discrimination of tree trunks from other objects such as traffic signs, street lamps, and other upright objects by utilizing both the cross- and co-polarized \ac{sar} images to identify foliage and tree trunks respectively. These features could then ultimately be used to then perform radar-image-based odometry and localization.

While the presented images show early results of a low frequency automotive \ac{sar} system, improvements could be made to increase the performance of the system. The presented system uses a monostatic approximation of the transmit and receive antennas, however in practice the antennas are positioned at either end of the vehicle's roof rack to minimize mutual coupling. This has the effect of modifying the parabolic shape traced by the range to a target as the platform passes a point scatterer and will degrade the image focusing when a monostatic approximation is used, especially near the vehicle; a bistatic formulation taking into account the spacing between the antennas would improve the focusing in this regard. The system shown in this paper also operates on the assumption of flat topography with a perfectly linear and constant platform velocity; naturally, these are not valid for most driving circumstances. \hlb{When the assumption of linear, constant velocity trajectory is violated, error manifests between the modeled and actual $x_n$ and $y_n$ terms (the latter of which was assumed zero in Section \ref{image-formation} due to the motion model). Due to this non-uniform sample, the cross-range Fourier transform will contain errors in the wavenumber $k_x$. Similarly, when performing the Stolt interpolation the estimation of the mapping from $(k_x,\,k_r)\rightarrow k_y$ will contain errors due to both the along-track and cross-track deviations of the platform being neglected. The cause of the measured deviation arises from two phenomena:
\begin{itemize}
	\item[1)] known deviations from the planned trajectory due to vehicle drift and vibration that is measured via \ac{gnss}, \ac{imu}, or other means, and
	\item[2)] unknown deviations from the planned trajectory due to uncertainty in the positioning sensing system.
\end{itemize} 
The type 1 deviations can be directly measured and compensated for via \acf{moco}, assuming the sensor uncertainty is lower than the magnitude of the deviations \cite{fornaro1999trajectory}\cite[Chapter 8.7]{richards2014fundamentals}. Several automotive \ac{sar} \ac{moco} techniques for non-linear trajectories have recently been demonstrated which perform phase compensation of the sampled \ac{sar} data prior to image formation by utilizing the kinematics measured on the platform during the \ac{sar} acquisition. An \ac{imu}-based motion compensation algorithm for automotive \ac{sar} is described and demonstrated in \cite{wu2011motion}, and more recently an \acp{imu} and \acp{gnss} automotive motion compensation algorithm was demonstrated in \cite{tagliaferri2021navigation}. Other's have demonstrated radar-only ego-motion estimation used for motion compensation \cite{iqbal2021imaging}. Type 2 deviations, on the other hand, are produced by unmeasurable error and thus cannot be directly corrected for, however many autofocusing algorithms exist to perform image sharpening on data with unknown errors \cite[Chapter 7]{jansing2021introduction}\cite[Chapter 8.7]{richards2014fundamentals}. One such algorithm is \textit{map drift} which uses a sub-aperture technique which divides the aperture into $N$ sub-sampled apertures, using every $n$th sample to create a sub-aperture; these are then cross-correlated and the time-delay offset is used to fit a phase error function to the trajectory which is used to apply a phase correction to the samples collected at each position in the synthetic aperture. A benefit of the autofocusing approaches is that it they only require the measured \ac{sar} data to perform the focusing, no auxiliary sensors, meaning that it may also be employed as a secondary focusing technique, after a kinematics-based motion compensation technique like those in \cite{wu2010novel, tagliaferri2021navigation, iqbal2021imaging} to address any error due to sensor inaccuracies.}

Finally, while low frequency automotive \ac{sar} has some clear advantages compared to the typical millimeter-wave bands, namely the ability to penetrate foliage and reduced atmospheric attenuation, the ability to reuse \ac{v2x} communications waveforms as the ranging mechanism would provide even greater incentive to consider the use of this frequency spectrum for automotive applications. The most common techniques recently proposed for \ac{jrc} perform some form of matched filtering on the transmitted communications waveform to determine the target range and relative Doppler shift \cite{reichardt2012demonstrating, ellison2020high, kumari2017ieee, dokhanchi2019mmwave, liu2020joint, duggal2020doppler}, however other techniques have considered time-domain multiplexing the radar and communications waveforms \cite{moghaddasi2013improved}. Most matched filtering based techniques focus on \acf{ofdm} waveforms as these are commonly employed in modern communications systems due to their high spectral efficiency and robustness to multipath channel distortion \cite{lathi2009modern}. However, due to the variable nature of the communications payloads, the ambiguity function of the waveform is constantly changing making it challenging to predict radar performance at any given time; to avoid this, other techniques have proposed using communication preambles which commonly include fixed training and channel estimation sequences \cite{ellison2020high, kumari2017ieee, liu2020joint, duggal2020doppler}. These fields can be used to obtain a consistent deterministic ranging waveform with predictable performance and relaxes the requirement for knowledge of the data being transmitted. While the benefit of reusing these signals is advantageous from a spectral interference reduction and hardware reuse viewpoint, the system performance is also critical for the adoption of \ac{jrc}. Currently, high throughput communication standards such as 802.11ac and 802.11ax implement channel bandwidths of up to \SI{160}{\mega\hertz} enabling range resolutions on the order of a meter, or potentially smaller through the use of non-adjacent channel bonding and the wider channel bandwidths proposed for future standards. The accuracy of a ranging waveform in a given scenario is highly dependent on the waveform shape which has many tradeoffs that can be made for ranging accuracy, Doppler tolerance, and side-lobe mitigation, however a recent study using a relatively low frequency, \SI{5.25}{\giga\hertz} 802.11ac legacy preamble as a ranging waveform experimentally achieved a ranging accuracy on the order of \SI{1.9}{\milli\meter} \cite{ellison2020high}, demonstrating the ability for existing 802.11 communications signals to be applicable for \ac{sar} applications in the \textit{C}-band. 

\hlb{One critical item which must be considered while implementing a \ac{jrc} using waveform reuse is the possibility of interference from outside transmitters communicating in the same frequency band.  This has the effect of decreasing the \acf{sinr} which in turn can significantly reduce detection range and create ghost targets in the radar image due to partial correlation between the transmitted signal and other signals in the environment \cite{brooker2007mutual, sit2014mimo}. In addition, due to the lower free-space loss of the \textit{C}-band compared with millimeter-wave bands commonly used by automotive radar, the possible interference range increases. While the randomness of data payloads can mitigate the potential for ghost targets in a matched filter, communications signals often utilize standardized payloads for certain parts of messages such as the training sequences and datagram headers. To some extent the ghost images produced via common packet features could be mitigated by using encrypted transmissions which, by design, should minimize the correlation between two packets (i.e., waveforms) containing the same data but encrypted using differing keys. Waveform orthogonality can also be coordinated in a cooperative manner by assigning transmitters in the network to transmit in an orthogonal manner (time, frequency, space, etc.) to reduce the number of interferers in the environment, as is commonly implemented in multi-user wireless communication channels \cite{lathi2009modern}.}

\hlb{One possible physical realization of this technique could include reusing existing \ac{v2x} transmitters, which are typically omnidirectional, to illuminate the scene around the vehicle. To receive the signals for proper \ac{sar} processing, the scatters should come from only one side of the vehicle's track to avoid ambiguities in scene reconstruction, thus, a minimum of two receive antennas, one on either side of the vehicle, would be required; ideally these antennas would have a wide beamwidth to provide fine imaging resolution as described in \eqref{cr_resolution}, and thus could be shared with \ac{v2x} reception. However, special considerations would need to be made for the transmit antenna design to achieve a fully polarimetric imaging system as the transmit antenna would need to be capable of transmitting both vertically and horizontally polarized waves in a controllable fashion in directions orthogonal to the vehicle track.  It should be noted, however, that the \ac{sar} concept can be implemented without the use of polarimetry and still produce scene scattering intensity images which may still be useful.}

\section{Conclusion}
\label{conclusion}

In this paper an experimental demonstration showing the efficacy of low frequency \ac{sar} for automotive use was presented, the unique foliage penetration aspects of the longer wavelength carrier were highlighted, and the benefits of using polarimetric \ac{sar} on a vehicle for foliage and landmark discrimination were also presented.  In addition to the results shown, combining the techniques employed in this paper with the recent advances in \ac{jrc} using low frequency \ac{v2x} signals will open a new regime for automotive radar sensing, enabling a once unused portion of the spectrum to support foliage penetrating detection and localization tasks. \hlb{While there is still significant work to be done before low-frequency \ac{jrc} automotive \ac{sar} systems can exist in a production environment, the technology would complement the currently existing modalities, such as millimeter-wave radar and lidar, adding the the multispectral array of sensors on \acp{av} to enable higher probability of detection and classification while not adding extra energy to the already active \textit{C}-band spectrum.}

\begin{IEEEbiography}[{\includegraphics[width=1in,height=1.25in,clip,keepaspectratio]{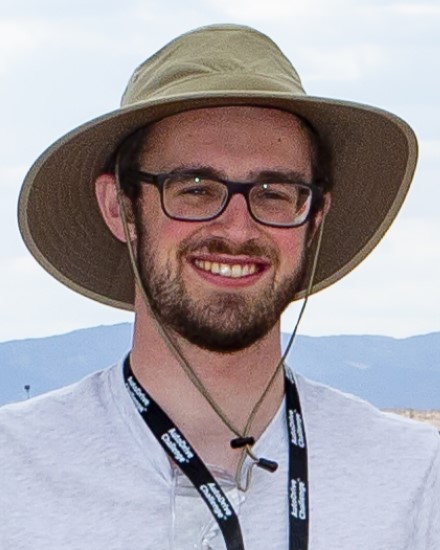}}]{Jason M. Merlo} received the B.S. degree in computer engineering from Michigan State University, East Lansing, MI, USA in 2018, where he is currently the Ph.D. degree in electrical engineering. From 2017-2021 he was project manager and electrical systems lead of the Michigan State University AutoDrive Challenge team. His current research interests include distributed radar, wireless synchronization, interferometric arrays, and automotive radar applications.
\end{IEEEbiography}
\vskip 0pt plus -1fil

\begin{IEEEbiography}[{\includegraphics[width=1in,height=1.25in,clip,keepaspectratio]{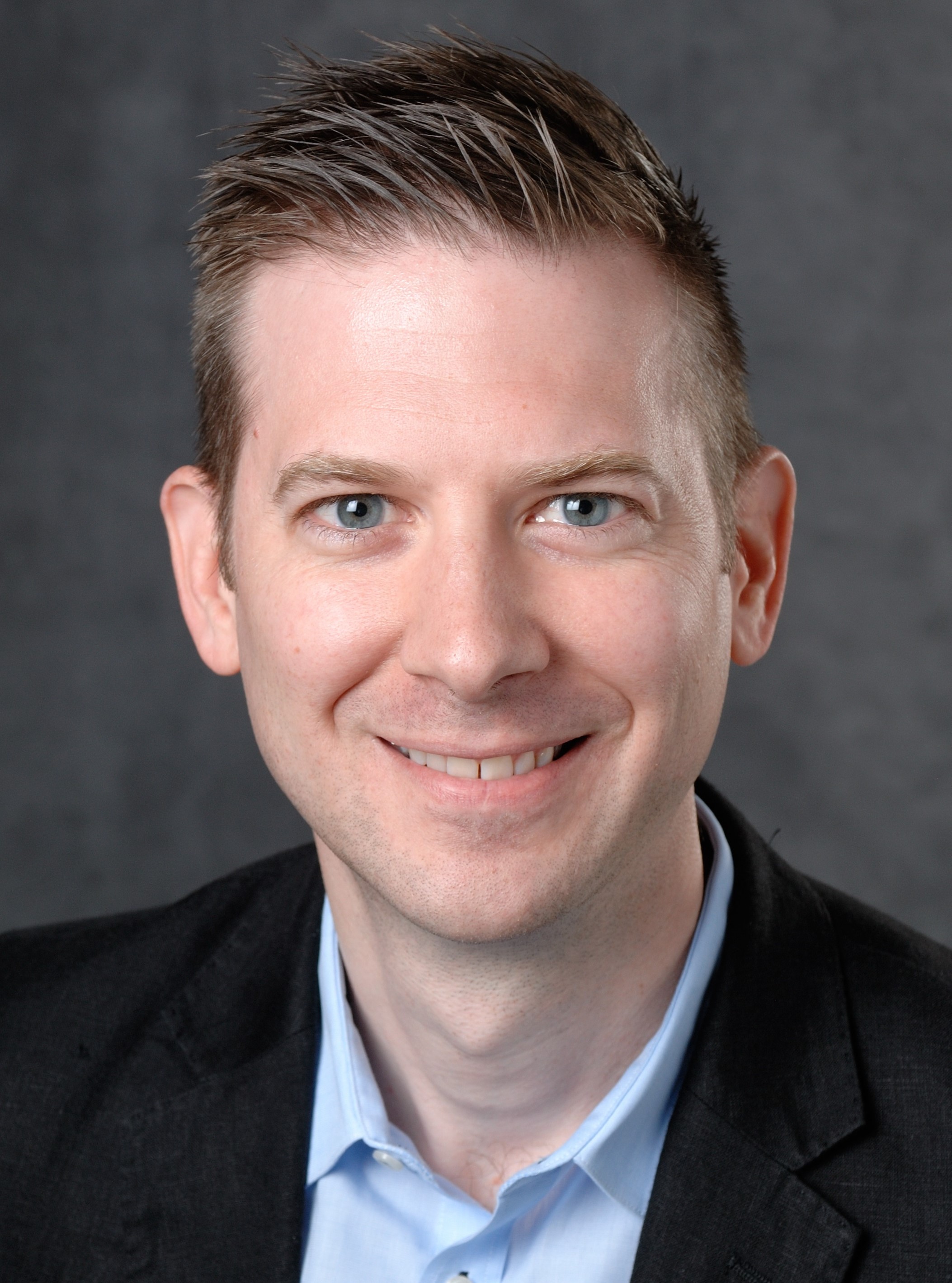}}]{Jeffrey A. Nanzer} (S'02--M'08--SM'14) received the B.S. degrees in electrical engineering and in computer engineering from Michigan State University, East Lansing, MI, USA, in 2003, and the M.S. and Ph.D. degrees in electrical engineering from The University of Texas at Austin, Austin, TX, USA, in 2005 and 2008, respectively.

From 2008 to 2009, he was a Post-Doctoral Fellow with Applied Research Laboratories, University of Texas at Austin, where he was involved in designing electrically small HF antennas and communication systems. From 2009 to 2016, he was with The Johns Hopkins University Applied Physics Laboratory, Laurel, MD, USA, where he created and led the Advanced Microwave and Millimeter-Wave Technology Section. In 2016, he joined the Department of Electrical and Computer Engineering, Michigan State University, where he held the Dennis P. Nyquist Assistant Professorship. He is currently an Associate Professor. He has authored or co-authored more than 175 refereed journal and conference papers, authored the book Microwave and Millimeter-Wave Remote Sensing for Security Applications (Artech House, 2012), and co-authored book chapters in Short-Range Micro-Motion Sensing with Radar Technology (IET Press, 2019) and Wireless Transceiver Circuits (Taylor \& Francis, 2015). His current research interests include distributed arrays, radar and remote sensing, antennas, electromagnetics, and microwave photonics.

Dr. Nanzer is a member of the IEEE Antennas and Propagation Society Education Committee and the USNC/URSI Commission B. He was a founding member and the First Treasurer of the IEEE APS/MTT-S Central Texas Chapter. He served as the Vice Chair for the IEEE Antenna Standards Committee from 2013 to 2015. He was the Chair of the Microwave Systems Technical Committee (MTT-16), IEEE Microwave Theory and Techniques Society from 2016 to 2018. He was a recipient of the Outstanding Young Engineer Award from the IEEE Microwave Theory and Techniques Society in 2019, the DARPA Director’s Fellowship in 2019, the National Science Foundation (NSF) CAREER Award in 2018, the DARPA Young Faculty Award in 2017, and the JHU/APL Outstanding Professional Book Award in 2012.  He is currently an Associate Editor of the IEEE TRANSACTIONS ON ANTENNAS AND PROPAGATION.

\end{IEEEbiography}
\bibliography{automotive_sar_2021_v1.2_FINAL_arxiv.bib}
\bibliographystyle{IEEEtran}

\end{document}